\documentclass[aps,pre,onecolumn]{revtex4}
\usepackage{graphicx, bm, bbm,amsmath, amssymb, epsfig}

\pdfoutput=1

\newcommand{\nn}{\noindent}

\newcommand{\bq}{\begin{align}}
\newcommand{\eq}{\end{align}}
\newcommand{\etal}{{\it{et al.}}~}

\begin{document}
\title{Efficient sliding locomotion with isotropic friction}
\author{Silas Alben}
\affiliation{Department of Mathematics, University of Michigan,
Ann Arbor, MI 48109, USA}
\email{alben@umich.edu}

\date{\today}

\begin{abstract}
Snakes' bodies are covered in scales that make it easier to slide in some directions than in others.
This frictional anisotropy allows for sliding locomotion with an undulatory gait, one of the most common
for snakes. Isotropic friction is a simpler situation (that arises with 
snake robots for example) but is less understood. 
In this work we regularize a model for sliding locomotion to allow for static friction.
We then propose 
a robust iterative numerical method to study the efficiency
of a wide range of motions under isotropic Coulomb friction. We find that simple undulatory
motions give little net locomotion in the isotropic regime. We compute 
general time-harmonic motions of three-link bodies and find three local optima for efficiency. The
top two involve static friction to some extent. We then propose a class of smooth body motions 
that have similarities to concertina locomotion (including the involvement of static friction) and can 
achieve optimal efficiency for both isotropic and anisotropic friction. 
\end{abstract}


\maketitle

\section{Introduction \label{sec:Introduction}}

Snake locomotion has attracted the interest of biologists and
engineers for several decades 
\cite{gray1946mechanism,gray1950kinetics,jayne1986kinematics,socha2002kinematics,HaCh2010b,MaHu2012a}. 
Many locomoting animals use appendages such as legs, wings, or fins
to exert a force on the substrate or surrounding fluid, and
propel the rest of the body forward \cite{dickinson2000animals}.
Snakes lack appendages, and thus it is less clear which parts of the snake
body should exert propulsive forces, and at which instants during the
motion, to move forward efficiently. 

A typical way to understand how organisms move 
is to study physical or computational models and compare their 
motions with those of the actual organisms 
\cite{dickinson2000animals,hohenegger2010stability,olson2011coupling,lim2012fluid,jones2016bristles}. 
One can take a step further and
pose and solve optimization problems for the models. This can
suggest locomotion strategies that are effective for man-made
vehicles \cite{bar2005biomimetics,jakimovski2011biologically,roper2011review}. 
It can also help understand why organisms have evolved in particular
ways under a multitude of 
constraints \cite{jacob1977evolution,alexander1996optima,langerhans2010ecology}.

Often what is optimized is
a measure of the efficiency of locomotion. For example, one
can maximize the average speed for a given time-averaged
power expended by the
organism. One can study the effects of physical parameters and constraints 
by varying them and studying how the optimal solutions change.
Well-known examples are optimization studies of organisms
moving in low- \cite{BeKoSt2003a,AvGaKe2004a,TaHo2007a,fu2007theory,spagnolie2010optimal,crowdy2011two,bittner2018geometrically}
and high-Reynolds-number fluid flows
\cite{lighthill1975mathematica,childress1981mechanics,sparenberg1994hydrodynamic,alben2009passive,michelin2009resonance,peng2012bb,gazzola2015gait}.
For locomotion in frictional (terrestrial or granular) media, frictional forces can result in distinctive
modes of efficient (or optimal) locomotion \cite{GuMa2008a,aguilar2016review}.

Snakes are limbless reptiles with elongated bodies, supported by
a backbone with 100--500 bony segments (vertebrae) 
\cite{lillywhite2014snakes}. The vertebrae allow for high
flexibility particularly in the lateral (side-to-side) 
direction, with less flexibility for 
vertical (dorso-ventral) bending or for torsion.
Running along the backbone are muscles that
attach to the sides of the vertebrae and cause bending. 
The snake body
is covered in a skin with a compliance (stretchability)
greater than that of mammalian skin, and widely
variable across species \cite{jayne1988mechanical}. 
The outside of the skin is covered in hardened, keratinous 
scales.
Scales on the belly are arranged so that friction
is lower when the snake slides towards its head and higher 
when it slides towards its tail. Muscles attach to scales
on the belly and can raise and lower them, modulating
their frictional properties and providing a 
gripping ability \cite{seigel1987snakes}.

On the basis of experiments and modeling, Hu and Shelley wrote
that ``snake propulsion on flat ground, and possibly in general, 
relies critically on the frictional anisotropy of their scales'' and measured the
friction coefficients for snake specimens sliding in different directions: 
$\mu_f$ (for a snake sliding forward, towards the head),
$\mu_b$ (sliding backward, towards the tail), and $\mu_t$ (sliding transverse to 
the body axis)
\cite{HuNiScSh2009a}. 
It is difficult to measure friction coefficients for moving snakes 
because their direction of motion and friction coefficients usually vary over their bodies.
Hu and Shelley found $\mu_b \approx 1.3 \mu_f$ and 
$\mu_t \approx 1.7 \mu_f$ for corn and milk snakes on cloth
\cite{HuSh2012a}.
Marvi and Hu measured forward and backward friction coefficients of
corn snakes by placing them on styrofoam inclines
and allowing them to slide head-first and tail-first under gravity \cite{MaHu2012a}. They found 
$\mu_b \approx 1.6 \mu_f$, and that conscious snakes' friction
coefficients are about twice those of unconscious snakes, which were
the focus of previous snake scale friction measurements \cite{gray1950kinetics,HuNiScSh2009a}. 
When conscious, snakes can increase the angles of their scales to 
grip the surface, increasing friction. 
 Hu and Shelley also studied the motions of snakes
wearing cloth sleeves, so that the scales do not contact the
substrate, giving a representation of isotropic friction ($\mu_t = \mu_b = \mu_f$). 
They found that when the snakes undulate while wearing a sleeve, there is little if
any forward motion \cite{HuNiScSh2009a,goldman2010wiggling}.

Transeth \etal used experiments and simulations to
show that for lateral undulation with isotropic friction,
locomotion is possible but slow without barriers to
push against \cite{transeth2008snake,transeth2009survey}.
Others have found that snake robots can achieve locomotion 
with isotropic friction using 3D motions:
sinus-lifting (slightly lifting the peaks of the body wave curve off the ground during
lateral undulation),
sidewinding, inchworm motions, and lateral rolling \cite{ohno2001design,liljeback2012review}.
Chernousko simulated particular gaits of multilinked
bodies with various friction coefficients and
found that locomotion could be obtained
with isotropic friction \cite{chernousko2005modelling}.
Wagner and Lauga studied the locomotion of a two-mass system moving
in one dimension with isotropic friction (equal in the 
forward and backward directions) and found
that locomotion is possible if the two masses have different
friction coefficients and the length of the link connecting them has
an asymmetric stroke cycle \cite{wagner2013crawling}. 
For the swimming of microorganisms in a viscous fluid (at zero Reynolds number),
the drag anisotropy of long slender bodies and appendages is known
to be essential for locomotion \cite{lauga2009hydrodynamics}.

\begin{figure} [h]
           \begin{center}
           \begin{tabular}{c}
               \includegraphics[width=6.5in]{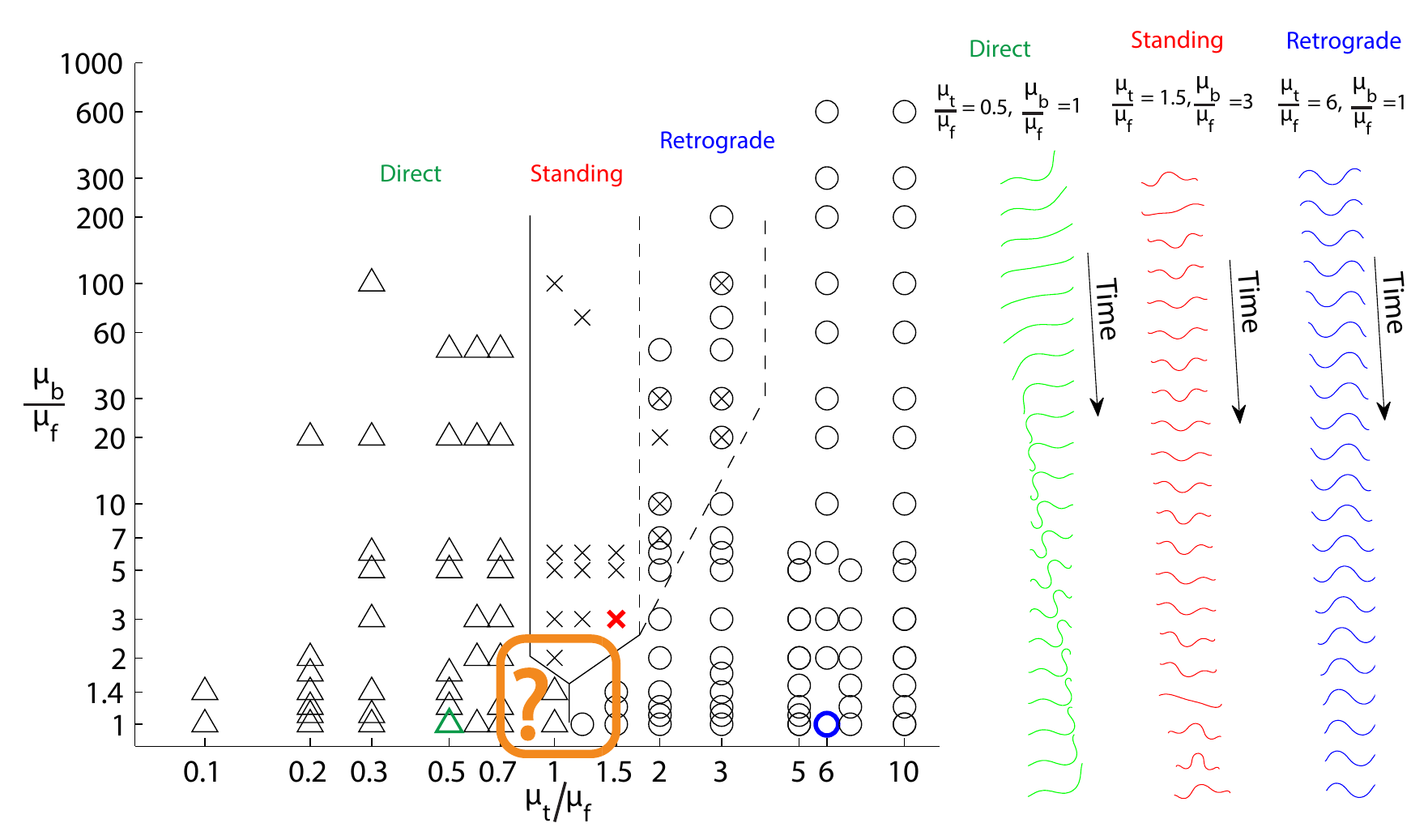} \\
           \vspace{-.25in} \hspace{-.25in}
           \end{tabular}
          \caption{\footnotesize Left: Classification of local optima across friction
coefficient space, computed in \cite{AlbenSnake2013}. Right: Three sequences of snapshots of locally optimal motions giving
examples of direct, standing, and retrograde waves. These occur at particular friction coefficient ratios, listed above the snapshots
and marked with green, red, and blue symbols in the panel at left. The three sequences of snapshots are given
over one period of motion, and displaced vertically to
enhance visibility but with the actual horizontal displacement.
 \label{fig:PhasePlaneMotionsIsotropicFig}}
           \end{center}
         \vspace{-.10in}
        \end{figure}

In a previous theoretical/computational study we optimized smooth snake body kinematics for efficiency, starting
from random initial ensembles \cite{AlbenSnake2013}. 
The kinematics were described by the coefficients of a double series, Fourier in time 
(with unit period) and Chebyshev (polynomials) in arc length along the
body axis, truncated at 45 modes (9 temporal by 5 spatial) and in some cases 190 modes
(19 temporal by 10 spatial). The searches were begun at random points in the
45- and 190-dimensional spaces of these coefficients. 
We searched for smooth time-periodic body kinematics
that maximize a definition of efficiency---the net distance traveled in one period divided by
the work done against friction in one period \cite{AlbenSnake2013}. The optimizers were calculated and 
classified as shown in figure \ref{fig:PhasePlaneMotionsIsotropicFig}, 
across the space of $\mu_t/\mu_f$ (horizontal axis) and 
$\mu_b/\mu_f$ (vertical axis).
Many of the local optima could be classified
as retrograde traveling waves---waves of curvature moving opposite to the body's direction
of motion (i.e. lateral undulation)---prevalent for 
$\mu_t/\mu_f \gtrsim 6$; symmetric standing waves, observed for
$\mu_b/\mu_f \geq 2$ and $0.7 < \mu_t/\mu_f \leq 3$; or direct waves---waves of 
curvature moving with the body's direction of motion---observed
for $\mu_t/\mu_f \lesssim 0.7$. Direct waves have also been observed in the 
undulatory swimming of polychaete worms, with appendages extending perpendicular
to the body axis \cite{taylor1952analysis,sfakiotakis2009undulatory}. 
Examples of these three classes of optima are shown in the snapshots
on the right side of figure \ref{fig:PhasePlaneMotionsIsotropicFig}. In
this study, one possible local optimum was observed with isotropic friction
$\mu_b/\mu_f = \mu_t/\mu_f = 1$, but the efficiency gradient norm was only reduced by about
two orders of magnitude from the random initial kinematics \cite{AlbenSnake2013}. 
Usually computations did not converge to local
optima in the vicinity of isotropic friction
(orange box in figure \ref{fig:PhasePlaneMotionsIsotropicFig}). Because isotropic
friction is common for snake robots (e.g. without scales) \cite{liljeback2012review}, is close to the measured friction
coefficients for real snakes \cite{HuSh2012a}, and is physically the simplest situation, a better
understanding of planar locomotion in this regime is useful. Isotropic friction is also a 
model of situations where snake scales are less effective---e.g. on loose, sandy, or slippery terrain
\cite{maladen2011undulatory}.
Effective kinematics
for planar locomotion with isotropic friction is the main topic of this study.





\section{Model \label{sec:Model}}

\begin{figure} [h]
           \begin{center}
           \begin{tabular}{c}
               \includegraphics[width=4in]{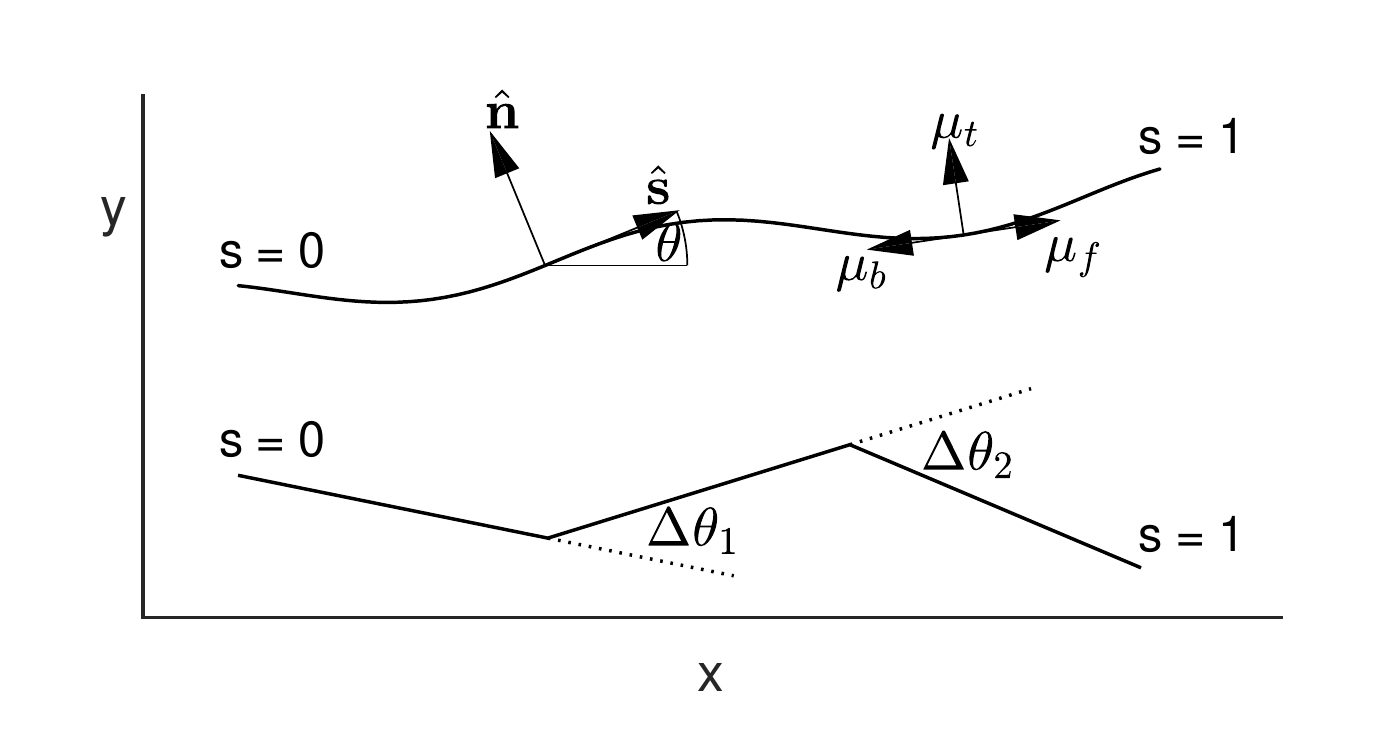} \\
           \vspace{-.25in}
           \end{tabular}
          \caption{\footnotesize Schematic diagrams of model snakes. Top: a smooth planar
curve
parametrized by arc length $s$ (nondimensionalized by snake length),
 at an instant in time.  The tangent angle
and the unit vectors
tangent and normal to the curve at a point are labeled. Vectors representing
forward, backward and transverse velocities are shown with
the corresponding friction coefficients $\mu_f$, $\mu_b$, and $\mu_t$.
Bottom: a three-link snake with changes in angles $\Delta\theta_1$ (here positive) and
$\Delta\theta_2$ (here negative) between the links.
 \label{fig:SmoothAnd3LinkSchematic}}
           \end{center}
         \vspace{-.10in}
        \end{figure}

We use the same Coulomb-friction
snake model as \cite{HuNiScSh2009a,HuSh2012a,JiAl2013} and other
recent works. 
The snake body is thin compared to its length, so for simplicity 
we approximate its motion
by that of a planar curve $\mathbf{X}(s,t) = (x(s,t), y(s,t))$,
parametrized by arc length $s$ and varying with time $t$. Schematic
diagrams are shown in figure \ref{fig:SmoothAnd3LinkSchematic}.

The tangent angle is denoted $\theta(s,t)$ and satisfies 
$\partial_s x = \cos\theta$ and $\partial_s y = \sin\theta$.
The unit vectors
tangent and normal to the curve are $\hat{\mathbf{s}} = (\partial_s x, \partial_s y)$ 
and $\hat{\mathbf{n}} = (-\partial_s y, \partial_s x)$
respectively. 
 The basic problem is to prescribe the time-dependent
shape of the snake in order to obtain efficient locomotion. We consider both
smooth bodies (figure \ref{fig:SmoothAnd3LinkSchematic}, top), and
three-link
bodies (figure \ref{fig:SmoothAnd3LinkSchematic}, bottom). The latter
are described by $\Delta\theta_1$ and $\Delta\theta_2$, the differences
between the tangent angles of the adjacent links. 

We prescribe the body shape as $\Theta(s,t)$, the tangent angle in the ``body frame,'' 
defined as a frame that rotates and translates so that at every time the body tail ($s=0$) lies at
the origin in the body frame and the body has zero tangent angle at the tail ($\Theta(0,t) = 0$).
In the three-link case, 
$\Theta(s,t) = \Delta \theta_1(t) H(s-1/3) + \Delta \theta_2(t) H(s-2/3)$, where
$H$ is the Heaviside function.
For all bodies (smooth and three-link), 
the tangent angle in the physical (or lab) frame is obtained by adding $\theta_0(t)$, the actual tangent
angle at the tail, to $\Theta(s,t)$:
\begin{align}
\theta(s,t) &= \theta_0(t) + \Theta(s,t) \label{theta0}.
\end{align}
\nn The body position in the lab frame is then obtained by integration:
\begin{align}
x(s,t) & = x_0(t) + \int_0^s \cos \theta(s',t) ds', \label{x0}\\
y(s,t) &= y_0(t) + \int_0^s \sin \theta(s',t) ds'. \label{y0}
\end{align}
\nn The tail position $(x_0(t), y_0(t))$ and tangent angle $\theta_0(t)$
(or equivalently, $\dot{x}_0(t), \dot{y}_0(t)$, and $\dot{\theta}_0(t)$)
are determined by the force and torque
balance for the snake, i.e. Newton's second law:
\begin{align}
\int_0^L \rho \partial_{tt} x ds &= \int_0^L f_x ds, \label{fx0} \\
\int_0^L \rho \partial_{tt} y ds &= \int_0^L f_y ds, \label{fy0} \\
\int_0^L  \rho \mathbf{X}^\perp \cdot \partial_{tt} \mathbf{X} ds
&= \int_0^L \mathbf{X}^\perp \cdot \mathbf{f} ds. \label{torque0}
\end{align}
\nn Here  $L$ is the body length, $\rho$ is the body's mass per unit length, and
$\mathbf{X}^\perp = (-y,x)$. For simplicity, the body is assumed to be locally inextensible
so $L$ is constant in time.
$\mathbf{f}$ is the force per unit length on the snake
due to Coulomb friction with the ground:
\begin{align}
\mathbf{f}(s,t) &= -\rho g \mu_t
\left( \widehat{\partial_t{\mathbf{X}}}\cdot \hat{\mathbf{n}} \right)\hat{\mathbf{n}}
-\rho g \left(\mu_f H\left(\widehat{\partial_t{\mathbf{X}}}\cdot \hat{\mathbf{s}}\right)
+ \mu_b \left(1-H\left(\widehat{\partial_t{\mathbf{X}}}\cdot \hat{\mathbf{s}}\right)\right)\right)
\left( \widehat{\partial_t{\mathbf{X}}}\cdot \hat{\mathbf{s}} \right)\hat{\mathbf{s}}. \label{friction}
\end{align}
\nn Again $H$ is the Heaviside function and the hats denote
normalized vectors. When $\|\partial_t{\mathbf{X}}\| = 0$
we define $\widehat{\partial_t{\mathbf{X}}}$ to be $\mathbf{0}$.
According to (\ref{friction}) the snake experiences
friction with different coefficients for motions in different directions.
The frictional coefficients are $\mu_f$, $\mu_b$, and $\mu_t$ for motions
in the forward ($\hat{\mathbf{s}}$), backward ($-\hat{\mathbf{s}}$),
and transverse (i.e. normal, $\pm\hat{\mathbf{n}}$) directions, respectively. In
general the snake velocity at a given point has both tangential and
normal components, and the frictional force density has
components acting in each direction. A similar decomposition of force
into directional components
occurs for viscous fluid forces on slender bodies \cite{cox1970motion}.

We assume that the body shape $\Theta(s,t)$ is periodic in time with period $T$,
similar to the steady locomotion of real snakes \cite{HuNiScSh2009a}.
We nondimensionalize equations (\ref{fx0})--(\ref{torque0}) by dividing
lengths by the snake length $L$, time by $T$, and mass by $\rho L$. Dividing
both sides by $g$ we obtain:
\begin{align}
\frac{L}{gT^2} \int_0^1 \partial_{tt} x ds &= \int_0^1 f_x ds, \label{fxa} \\
\frac{L}{gT^2}\int_0^1 \partial_{tt} y ds &= \int_0^1 f_y ds, \label{fya} \\
\frac{L}{gT^2}\int_0^1 \mathbf{X}^\perp \cdot \partial_{tt} \mathbf{X} ds
&= \int_0^1 \mathbf{X}^\perp \cdot \mathbf{f} ds. \label{torquea}
\end{align}
\nn In (\ref{fxa})--(\ref{torquea}) and from now on, all variables are
dimensionless. If the body accelerations are not very large, as is often the case
for robotic and real snakes \cite{HuNiScSh2009a}, $L/gT^2 \ll 1$,
which means that the body's inertia is negligible. By setting
inertia---and the left hand sides of (\ref{fxa})--(\ref{torquea})---to zero, we simplify the equations
considerably:
\begin{align}
\int_0^1 f_x ds &= \int_0^1 f_y ds = \int_0^1 \mathbf{X}^\perp \cdot \mathbf{f} ds = 0. \label{ftb}
\end{align}
\nn In (\ref{ftb}), the dimensionless force $\mathbf{f}$ is
\begin{align}
\mathbf{f}(s,t) = -\mu_t
\left( \widehat{\partial_t{\mathbf{X}}}\cdot \hat{\mathbf{n}} \right)\hat{\mathbf{n}}
- \left( \mu_f H(\widehat{\partial_t{\mathbf{X}}}\cdot \hat{\mathbf{s}})
+ \mu_b (1-H(\widehat{\partial_t{\mathbf{X}}}\cdot \hat{\mathbf{s}}))\right)
\left( \widehat{\partial_t{\mathbf{X}}}\cdot \hat{\mathbf{s}} \right)\hat{\mathbf{s}}. \label{friction1}
\end{align}
\nn  Similar models were used in 
\cite{GuMa2008a,HuNiScSh2009a,HuSh2012a,JiAl2013,AlbenSnake2013,wang2014optimizing,wang2018dynamics},
and the same model was found to agree well with the motions of biological
snakes in \cite{HuNiScSh2009a}.

\begin{figure} [h]
           \begin{center}
           \begin{tabular}{c}
               \includegraphics[width=6in]{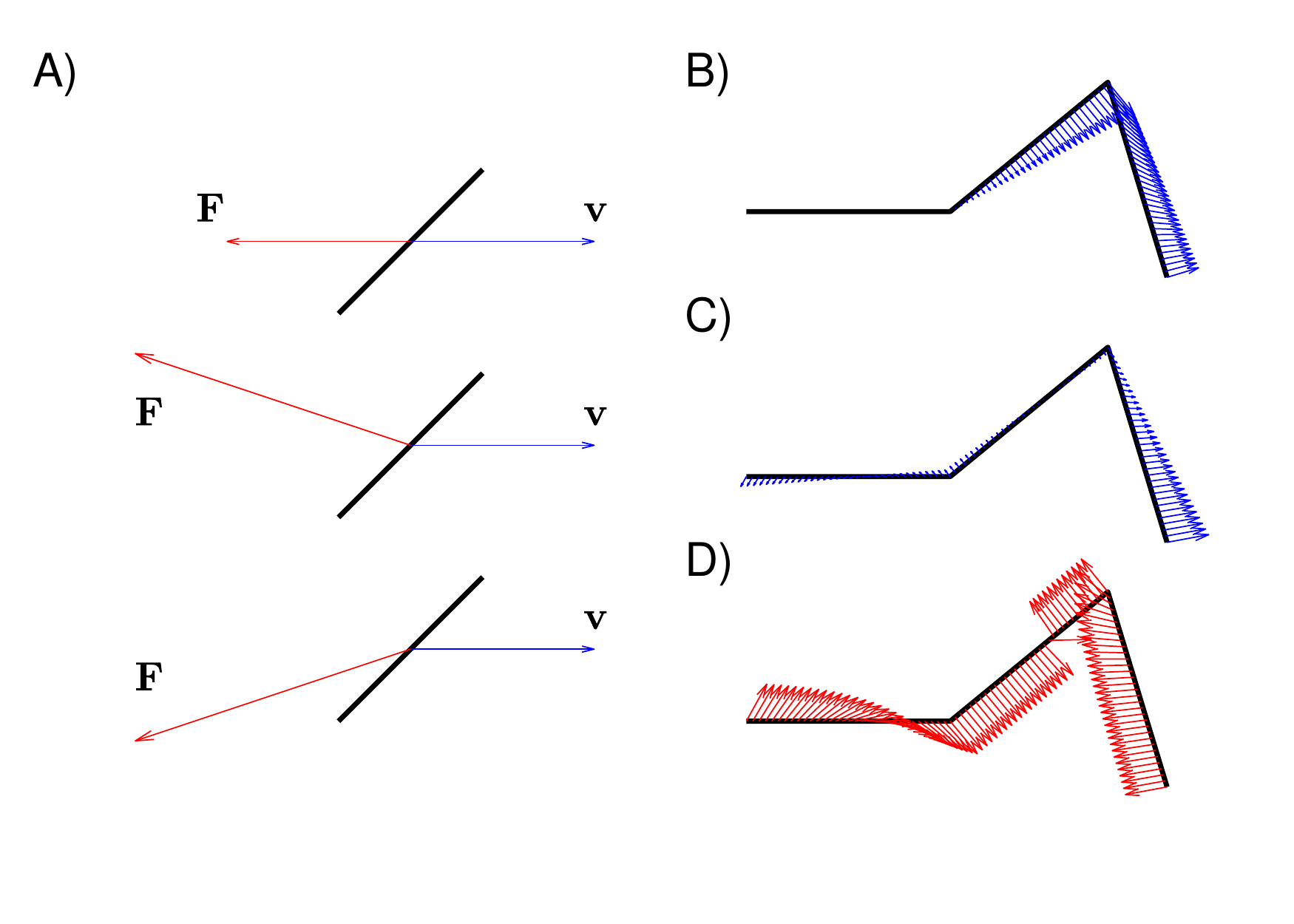} \\
           \vspace{-.25in}
           \end{tabular}
          \caption{\footnotesize A) Frictional force $\mathbf{F}$ (red vector) 
acting on a flat plate moving uniformly with horizontal velocity $\mathbf{v}$ (blue vector)
when the transverse friction coefficient $\mu_t$ is equal to (top), greater than (middle), or less than
(bottom) the forward friction coefficient $\mu_f$. B) Velocity distribution (blue vectors) on
a three-link body with zero translational and rotational velocities ($\dot{\mathbf{X}}_0$ and $\dot{\theta}_0$)
at the tail (left endpoint). Here
$\Delta\theta_1$ = 39 degrees, $\Delta\theta_2$ = -113 degrees, $\dot{\Delta\theta}_1$ =
-0.56, and $\dot{\Delta\theta}_2 = 1$.
C) Velocity distribution in the lab frame: the translational and rotational velocities at the tail 
are such that
the integrated force and torque due to the frictional force distributions (red vectors shown in panel D) are zero.
 \label{fig:FrictionForceDiagram}}
           \end{center}
         \vspace{-.10in}
        \end{figure}

Figure \ref{fig:FrictionForceDiagram} shows examples of the force-velocity relationship
expressed by (\ref{friction1}). Panel A shows the total frictional force $\mathbf{F}$ (red vector) on a flat plate with a 45-degree tangent angle
and uniform horizontal velocity $\mathbf{v}$ (blue vector) for three different choices of friction coefficients. At the top is isotropic friction, $\mu_f = \mu_t = 1$ ($\mu_b$ is not involved here since 
$\mathbf{v}\cdot \hat{\mathbf{s}} > 0$). With isotropic friction, $\mathbf{f}$ is
directed opposite to $\mathbf{v}$. The middle case has
$\mu_f = 1$ and $\mu_t = 2$, increasing the force component in the 
$\hat{\mathbf{n}}$-direction. The bottom case has instead $\mu_f = 2$ and $\mu_t = 1$, 
increasing the force component in the -$\hat{\mathbf{s}}$-direction.
 Panel B shows an example of a motion of a three-link body where the tail velocities
$\dot{x}_0(t), \dot{y}_0(t)$, and $\dot{\theta}_0(t)$ are zero. Here $\Delta\theta_1$ is 
decreasing and $\Delta\theta_2$ is increasing in time, resulting in the nonuniform velocity distribution
(piecewise linear in $s$) shown by the blue vectors. The force and torque balance equations are not
satisfied by this motion. Panel C shows the same motion but with $\dot{x}_0(t), \dot{y}_0(t)$, and 
$\dot{\theta}_0(t)$ chosen to satisfy equations (\ref{ftb}). This adds a counterclockwise rotation and
downward and leftward translation to the body. The resulting force distribution is shown by
the red vectors in panel D. The net force and torque from this distribution are zero. 
Although the velocities are small on the first two links, the forces are large---the
normalization of velocities in (\ref{friction1}) means that small velocities can give rise to $O$(1) forces.
The motion in panel C is approximately one in which only the third link is moving, rotating counterclockwise, but 
the small but nonzero velocities on the first two links are enough to give forces and torques that
balance those on the third link.

Instead of solving (\ref{ftb}) for $\{{x}_0(t), {y}_0(t), {\theta}_0(t)\}$ directly, we solve them
for $\{\dot{x}_0(t), \dot{y}_0(t), \dot{\theta}_0(t)\}$, which can be done (mostly) in parallel, 
speeding up the computations. We 
take time derivatives of (\ref{theta0})-(\ref{y0}), using vector notation for position:
\begin{align}
\partial_t \theta(s,t) &= \dot{\theta}_0(t) + \partial_t \Theta(s,t) \label{dttheta0}. \\
\partial_t \mathbf{X}(s,t) & = \dot{\mathbf{X}}_0(t) + \int_0^s  \left(\dot{\theta}_0(t) + \partial_t \Theta(s,t)\right)\hat{\mathbf{n}} ds'. \label{dtz0}
\end{align}
\nn Given $\Theta(s,t)$ and $\partial_t \Theta(s,t)$, we first solve (\ref{ftb}) with $\theta_0(t) = 0$ and 
$\mathbf{X}_0(t) = 0$ to obtain a solution $\{\dot{\mathbf{X}}_{0b}(t),\dot{\theta}_{0b}(t)\}$ 
in the body frame for the unknowns $\{\dot{\mathbf{X}}_0(t),\dot{\theta}_0(t)\}$ in
(\ref{dttheta0})-(\ref{dtz0}). 
The solution 
$\{\dot{\mathbf{X}}_{0b}(t),\dot{\theta}_{0b}(t)\}$ represents the tail velocity if the body is rotated by $-\theta_0(t)$
so that the tail has zero tangent angle. The position $\mathbf{X}$ and
tangent and normal vectors $\hat{\mathbf{s}}, \hat{\mathbf{n}}$
in the lab frame are simply those in the body frame rotated
by $\theta_0(t)$. If we set $\dot{\theta}_0(t) = \dot{\theta}_{0b}(t)$ and let
$\dot{\mathbf{X}}_0(t)$ be $\dot{\mathbf{X}}_{0b}(t)$ rotated by
$\theta_0(t)$, then we find that the lab frame velocity $\partial_t{\mathbf{X}}$ in
(\ref{dtz0}) is the body frame velocity rotated by $\theta_0(t)$. Hence $\mathbf{f}$ in 
(\ref{friction1}) is that in the body frame rotated by $\theta_0(t)$ and
$\mathbf{X}^\perp \cdot \mathbf{f}$ is unchanged (this dot product and those in
$\mathbf{f}$ are unchanged by the rotation)---so both 
$\mathbf{f}$ and $\mathbf{X}^\perp \cdot \mathbf{f}$ still integrate
to zero under the transformation from
the body to lab frame. To summarize: 
if $\{\dot{\mathbf{X}}_{0b}(t),\dot{\theta}_{0b}(t)\}$
solve (\ref{ftb}) with $\{\mathbf{X}_0(t), {\theta}_0(t)\}$ equal to zero (i.e. 
in the body frame),
then $\dot{\mathbf{X}}_0(t) = \mathbf{R}_{\theta_0(t)}\dot{\mathbf{X}}_{0b}(t)$
and $\dot{\theta}_0(t) = \dot{\theta}_{0b}(t)$
solve (\ref{ftb}) with general $\{\mathbf{X}_0(t), {\theta}_0(t)\}$, when the
body is also rotated by $\theta_0(t)$ (i.e. the body is in the lab frame). Here
\begin{align}
\mathbf{R}_{\theta_0(t)} =  \begin{pmatrix}
\cos \theta_0(t) & \hfill -\sin \theta_0(t)   \cr
\sin \theta_0(t) & \hfill \cos \theta_0(t) 
\end{pmatrix},
\end{align}
\nn the matrix that rotates by $\theta_0(t)$. 

We can solve for $\{\dot{\mathbf{X}}_{0b}(t),\dot{\theta}_{0b}(t)\}$
at all time steps in parallel, since only $\Theta(s,t)$ and $\partial_t \Theta(s,t)$ are required.
Then we integrate $\dot{\theta}_{0}(t) =  \dot{\theta}_{0b}(t)$ forward in time
to obtain the tail tangent angle starting from $\theta_0(0) = 0$ (an arbitrary constant
that sets the overall trajectory direction). Then we integrate
$\dot{\mathbf{X}}_0(t) = \mathbf{R}_{\theta_0(t)}\dot{\mathbf{X}}_{0b}(t)$
forward in time starting from $\mathbf{X}_0(0) = 0$ (another arbitrary constant)
to obtain the tail position in time. Then the complete body motion is known from
(\ref{theta0})-(\ref{y0}).

%
%
%
%
%
%
%
%

In this work
we will consider only motions that involve zero net rotation over one period,
i.e. $\theta_0(1) = \theta_0(0)$. Then the motion after one period is a pure
translation, with all points on the body moving the same distance
\begin{align}
d = \sqrt{\left(x_0(1) - x_0(0)\right)^2 +
\left(y_0(1) - y_0(0)\right)^2}. \label{dist}
\end{align}
The work done by the snake against friction over one period is
\begin{align}
W &= \int_0^1 \int_0^1 -\mathbf{f}(s,t) \cdot \partial_t\mathbf{X}(s,t)\, ds\, dt. \label{W}
\end{align}
When the body shape motion 
$\Theta(s,t)$ is uniformly sped up or slowed down---i.e. when 
\begin{align}
\Theta(s,t) \rightarrow \Theta(s,ct), \quad 
\partial_t \Theta(s,t) \rightarrow c\partial_t \Theta(s,ct)
\end{align}
\nn for some constant $c > 0$, then the force and torque balance equations are satisfied when
the tail motion undergoes the same scaling: 
\begin{align}
\{x_0(t), y_0(t), \theta_0(t)\}
\rightarrow \{{x}_0(ct), {y}_0(ct), {\theta}_0(ct)\}, \quad \{\dot{x}_0(t), \dot{y}_0(t), \dot{\theta}_0(t)\}
\rightarrow c\{\dot{x}_0(ct), \dot{y}_0(ct), \dot{\theta}_0(ct)\} \label{scaling}
\end{align}
\nn and so does the overall body motion:
\begin{align}
\mathbf{X}(s,t) \rightarrow \mathbf{X}(s,ct), \quad 
\partial_t \mathbf{X}(s,t) \rightarrow c\partial_t \mathbf{X}(s,ct).
\end{align}

\nn We can see this by first plugging the transformed quantities into (\ref{dttheta0})-(\ref{dtz0}),
to verify that those equations are still obeyed.
We also have $\widehat{\partial_t{\mathbf{X}}}(s, t) \rightarrow 
\widehat{\partial_t{\mathbf{X}}}(s, ct)$, and so
the frictional force $\mathbf{f}(s, t) \rightarrow \mathbf{f}(s, ct)$ by (\ref{friction1}),
assuming $c > 0$ (note
that $\hat{\mathbf{s}}(s, t) \rightarrow \hat{\mathbf{s}}(s, ct)$ and
$\hat{\mathbf{n}}(s, t) \rightarrow \hat{\mathbf{n}}(s, ct)$) and the
torque density $\mathbf{X}^\perp \cdot \mathbf{f}$ has the same transformation.
If $\mu_b = \mu_f$ then
the $H(\widehat{\partial_t{\mathbf{X}}}\cdot \hat{\mathbf{s}})$ term drops out of
$\mathbf{f}$ in (\ref{friction1}) and the same scaling holds for $c < 0$ also ($\mathbf{f}$ changes sign 
uniformly in this case). If instead $\mu_b \neq \mu_f$, then the solutions are not simply time-reversed when the shape change is time-reversed.

Imagine now that we take a given periodic motion and repeat it $n$ times in a period.
Then the velocities are multiplied by $n$, and so is the net distance $d$.
The same is true of $W$ since
in (\ref{W}), $\partial_t\mathbf{X}(s,t) \rightarrow n\partial_t\mathbf{X}(s,nt)$
and $\mathbf{f}$ is unchanged. Since
$d$ and $W$ both scale with the speed of the motion, it makes sense to define an efficiency as
\begin{align}
\lambda = \frac{d}{W}. \label{eta}
\end{align}
\nn which is the same when a given motion is sped up or slowed down. A somewhat more general problem, 
not pursued here, is
to find motions that maximize $d$ for a given $W > 0$, and then vary $W$. For small $W$,
only a limited
set of periodic motions---those with small amplitude---can perform work $W$ in a period. When
$W$ is large, large-amplitude motions can perform work $W$, but also small amplitude motions
by repeating the motion a given number of times. Hence as $W$ becomes larger we consider
a larger class of motions that can eventually approximate essentially any periodic motion.  

Next we will calculate $W$, $d$, and $\lambda$ for certain examples of motions (i.e. $\Theta(s,t)$)
with both isotropic ($\mu_f = \mu_t = \mu_b = 1$) and anisotropic friction.
Then we will focus on the isotropic case. We will examine the class of 
time-harmonic three-link motions and then propose a class of smooth motions that optimize
$\lambda$.

\begin{figure} [h]
           \begin{center}
           \begin{tabular}{c}
               \includegraphics[width=4in]{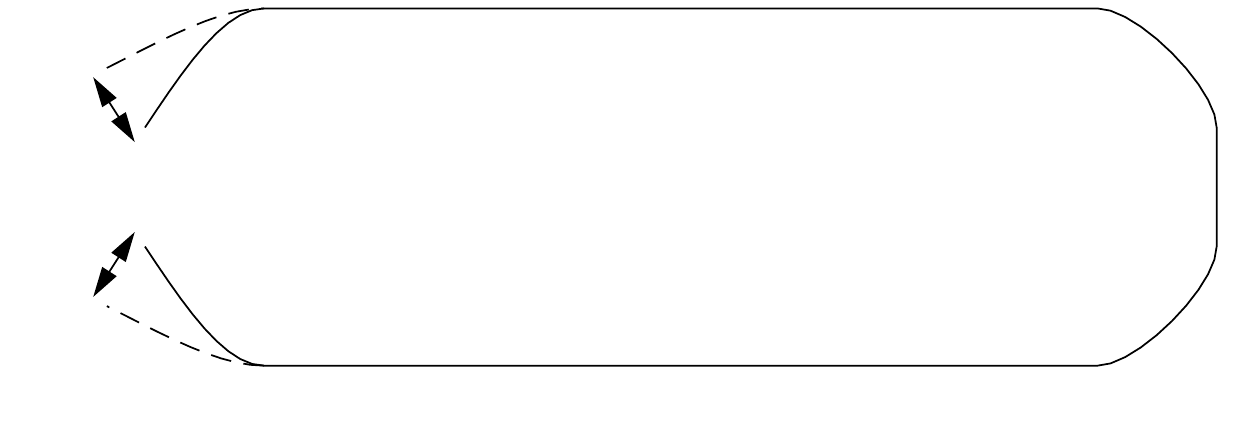} \\
           \vspace{-.25in} \hspace{-.25in}
           \end{tabular}
          \caption{\footnotesize Sketch of a body motion for which the kinetic friction model has no solution, so
a model of static friction is used.
 \label{fig:StaticFrictionSketch}}
           \end{center}
         \vspace{-.10in}
        \end{figure}

Equations (\ref{ftb}) assume
only kinetic friction is involved, but in reality there is also static friction.
In figure \ref{fig:StaticFrictionSketch} we show an example of a motion for which
the kinetic friction model has no solution.  That is, for the $\Theta(s,t)$ corresponding to this motion (not given mathematically here), 
no choice of $\{\dot{x}_0(t), \dot{y}_0(t), \dot{\theta}_0(t)\}$ can solve equations (\ref{ftb}).
Initially the body is given by the solid line. The two flaps on the left side oscillate periodically, sweeping out a region shown by arrows between the solid line and the dashed lines. On the upstroke, the combined vertical force and torque on the flaps from kinetic friction 
(\ref{friction1}) is zero by symmetry, but there is a net horizontal force to the right. If we assume isotropic friction, the horizontal force per unit length on the flaps from (\ref{friction1}) lies between 0 and 1, since the flaps move leftward and upward. The rest of the body cannot balance this force exactly for the following reasons.  Its motion can only be horizontal to maintain vertical force balance. Therefore,
by (\ref{friction1}) it has horizontal force per unit length -1, 0, or +1, and a much larger length than the flaps. None of these choices gives
zero net horizontal force on the body as a whole.
The problem is resolved physically by including static friction: a force density
between 0 and that given by kinetic friction when the velocity is zero \cite{bhushan2013introduction}. Further 
examples will be given (for three-link bodies) in section \ref{sec:3linkharm} (e.g. figure \ref{fig:StaticFrictionDiagram}).

To allow for static friction, we use a simple modification of (\ref{friction1}) involving a regularization parameter $\delta$:
\begin{align}
\mathbf{f}_\delta(s,t) &\equiv -\mu_t
\left( \widehat{\partial_t{\mathbf{X}}}_\delta\cdot \hat{\mathbf{n}} \right)\hat{\mathbf{n}}
- \left( \mu_f H(\widehat{\partial_t{\mathbf{X}}}_\delta\cdot \hat{\mathbf{s}})
+ \mu_b (1-H(\widehat{\partial_t{\mathbf{X}}}_\delta\cdot \hat{\mathbf{s}}))\right)
\left( \widehat{\partial_t{\mathbf{X}}}_\delta\cdot \hat{\mathbf{s}} \right)\hat{\mathbf{s}}, \label{frictiondelta} \\ 
\widehat{\partial_t{\mathbf{X}}}_\delta &\equiv \frac{\left(\partial_t x, \partial_t y\right)}{\sqrt{\partial_t x^2 +\partial_t y^2 + \delta^2}}. \label{delta}
\end{align}
\nn Here $\delta$ is small, $10^{-4}$ in our computations. We find empirically that there is little change
in the results (less than 1\% in relative magnitude) for $\delta$ in the range $(0, 10^{-4}]$. 
When $\sqrt{\partial_t x^2 +\partial_t y^2}$ is similar in magnitude to $\delta$, the force density in (\ref{frictiondelta}) 
varies between 0 and 1 in magnitude, times the appropriate friction coefficient. Therefore we obtain the full range of
force densities when velocities are very small, which approximates static friction. In addition to their simplicity, 
we find empirically that expressions
(\ref{frictiondelta})-(\ref{delta}) have desirable properties including the existence of unique solutions using
the numerical algorithm described next. More specifically, for all motions shown in the work, our iterative numerical method
(described next) finds a unique
solution $\{\dot{x}_0(t), \dot{y}_0(t), \dot{\theta}_0(t)\}$ to equations (\ref{ftb}) with $\mathbf{f}_\delta$
in place of $\mathbf{f}$, for a large number of initial guesses (covering a wide range including choices very far from the solution). Similar
types of Coulomb friction regularization (sometimes involving the arctangent function) have been used for many years in 
dynamical simulations involving friction \cite{popov2010contact,pennestri2016review}. In our case,
$\delta$ needs to be small compared to any physical velocities we wish to resolve. In particular, $\delta$
should be small compared to the speed of body deformations: 
the typical magnitude of $\partial_t\Theta(s,t)$ multiplied by the range of arc length in which it varies from zero.

\section{Numerical method}

In previous work \cite{AlbenSnake2013}, we computed solutions to equations (\ref{ftb}) using
quasi-Newton methods. Two major challenges of such methods are finding an 
initial guess that is sufficiently close for convergence, and choosing a step size in the line search
that moves the solution towards convergence. The components of $\mathbf{f}_\delta$ behave like
smoothed step functions near zero velocity. If the solution has velocities near zero 
(i.e. involves static friction), Newton's method requires a very good initial guess, within
$O(\delta)$ of the solution, to converge. The behavior is similar to that for the arctangent
function, a classic example used to illustrate the limited basin of attraction for Newton's method
near a root \cite{kelley1995iterative,dennis1996numerical}. 

To compute large numbers of solutions to (\ref{ftb}) in parallel, we have developed 
a more robust iterative 
scheme that converges with any initial guess (for all cases studied, a large number including those in
this work) and does not require a line search. The iteration is a fixed point iteration using
a linearization of the regularized version of equations (\ref{ftb}). At time $t$, given
$\Theta(s,t)$ and a guess $\{\dot{x}^n_0(t), \dot{y}^n_0(t), \dot{\theta}^n_0(t)\}$,
we use (\ref{dttheta0})-(\ref{dtz0}) to compute the corresponding 
$\{\partial_t{x}^n(s,t), \partial_t{y}^n(s,t), \partial_t{\theta}^n(s,t)\}$, and then solve
\begin{align}
\int_0^1 \tilde{f}_{\delta x} ds &= \int_0^1 \tilde{f}_{\delta y} ds = \int_0^1 \mathbf{X}^\perp \cdot \tilde{\mathbf{f}}_\delta ds = 0 \label{ftbtilde}
\end{align}
\nn for a new iterate $\{\dot{x}^{n+1}_0(t), \dot{y}^{n+1}_0(t), \dot{\theta}^{n+1}_0(t)\}$ where 
\begin{align}
\tilde{\mathbf{f}}_\delta(s,t) &\equiv -\mu_t
\left( \tilde{\widehat{\partial_t{\mathbf{X}}}}_\delta\cdot \hat{\mathbf{n}} \right)\hat{\mathbf{n}}
- \left( \mu_f H(\tilde{\widehat{\partial_t{\mathbf{X}}}}_\delta\cdot \hat{\mathbf{s}})
+ \mu_b (1-H(\tilde{\widehat{\partial_t{\mathbf{X}}}}_\delta\cdot \hat{\mathbf{s}}))\right)
\left( \tilde{\widehat{\partial_t{\mathbf{X}}}}_\delta\cdot \hat{\mathbf{s}} \right)\hat{\mathbf{s}}, \label{frictiondeltatilde} \\ 
\tilde{\widehat{\partial_t{\mathbf{X}}}}_\delta &\equiv \frac{\left(\partial_t x^{n+1}, \partial_t y^{n+1}\right)}{\sqrt{\left(\partial_t x^n\right)^2 +
\left(\partial_t y^n\right)^2 + \delta^2}}. \label{deltatilde}
\end{align}
 \nn Iterate $n$ is used in the denominator of (\ref{deltatilde}), so the new iterate $\{\dot{x}^{n+1}_0(t), \dot{y}^{n+1}_0(t), \dot{\theta}^{n+1}_0(t)\}$ appears only in the numerator, and (\ref{ftbtilde})-(\ref{deltatilde})
depend linearly on it (in the body frame, where  $\mathbf{X}, \hat{\mathbf{s}}$, and
$\hat{\mathbf{n}}$ are known). Hence we obtain the new iterate $\{\dot{x}^{n+1}_0(t), \dot{y}^{n+1}_0(t), \dot{\theta}^{n+1}_0(t)\}$ by solving 3-by-3 linear systems at
each $t$ (decoupled when solving in the body frame). We observe empirically that this approach sacrifices the 
quadratic or superlinear convergence of Newton-type methods for linear (geometric) convergence. In almost all cases the
convergence is quite fast, however. There are a small number of cases involving static friction where the rate of
geometric convergence is slower. However these cases are sufficiently few that even with more iterates, the cost of
obtaining convergence is small. The loss of superlinear convergence is relatively modest compared to 
the increased simplicity and robustness of the algorithm.


\section{Examples of motions}

\begin{figure} [h]
           \begin{center}
           \begin{tabular}{c}
               \includegraphics[width=6in]{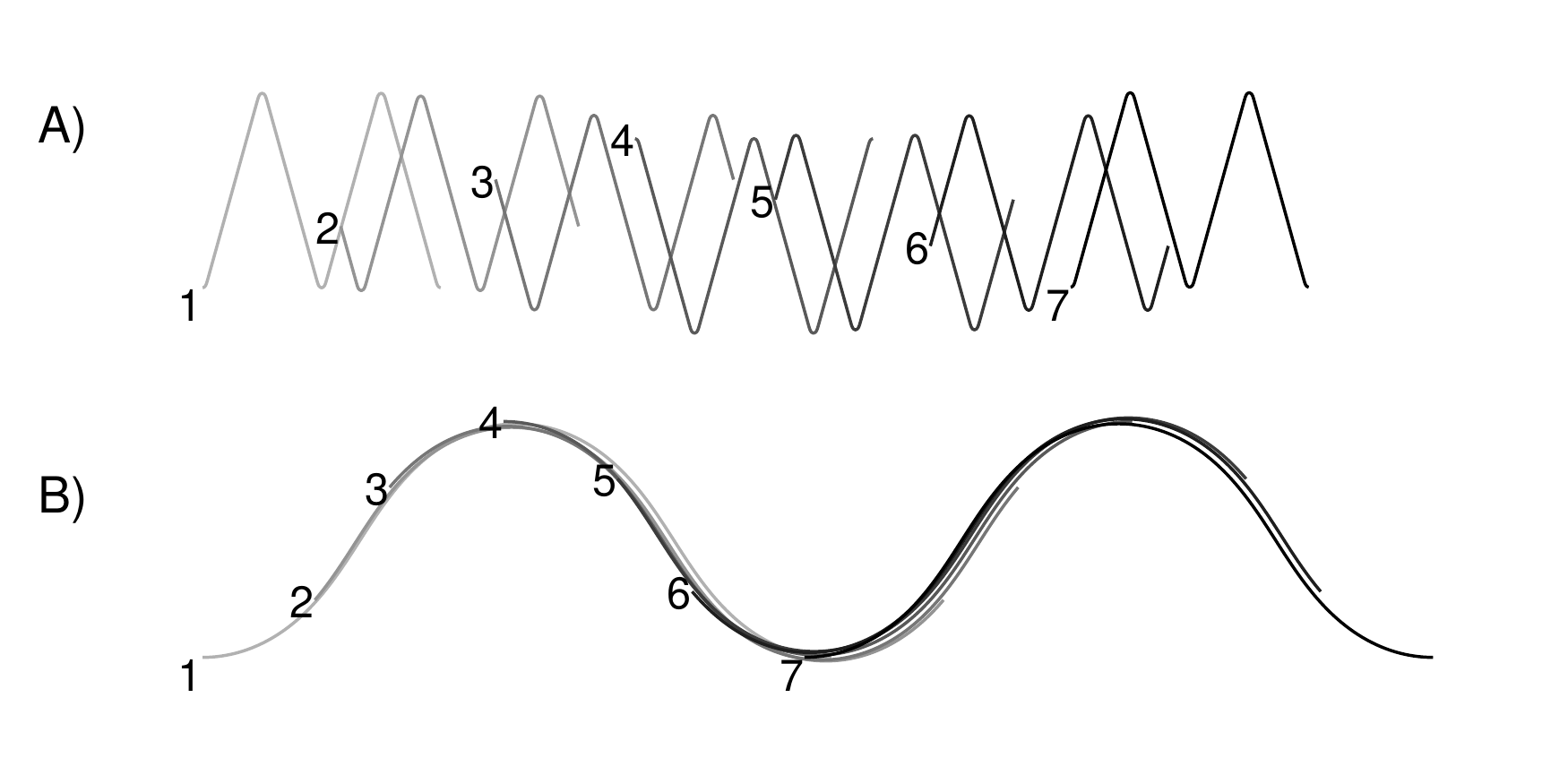} \\
           \vspace{-.25in} \hspace{-.25in}
           \end{tabular}
          \caption{\footnotesize Snapshots of the snake body 
when executing time-periodic traveling wave body deformations 
(over one time period, darker at later times, labeled near the tail in numerical order). Top: Rightward-moving
smoothed triangular deformation wave. Since $\mu_t = 1 \ll 100 = \mu_f$, the body moves
rightward (i.e. a direct wave).
The body tangent angle is
$\approx \pm 1.3$ in the straight regions ($\Theta(s,t) = 1.3\tanh(20\sin(2\pi(2s-t)))$). 
Bottom:  Leftward-moving sinusoidal deformation wave with wavelength 1
($\Theta(s,t) = \sin(2\pi(s+t))$). Since $\mu_f = 1 \ll 100 = \mu_t$, the body moves
rightward (i.e. a retrograde wave).
 \label{fig:OptimalAnisotropicMotionsFig}}
           \end{center}
         \vspace{-.10in}
        \end{figure}

We now present numerical solutions of the model described in section \ref{sec:Model}.
We show motions that are approximately optimal with very anisotropic friction, and then show how these motions perform with isotropic friction.

In figure \ref{fig:OptimalAnisotropicMotionsFig}A we show snapshots of the body when
executing a rightward-moving smoothed triangular wave ($\Theta(s,t) = 1.3\tanh(20\sin(2\pi(2s-t)))$) with friction much smaller in the transverse direction than in the tangential direction ($\mu_t = 1 \ll 100 = \mu_f = \mu_b$). The motion is almost entirely in the transverse direction,
and due to the almost vertical body slope, the transverse direction is approximately horizontal,
close to the direction of locomotion. Consequently the efficiency $\lambda$ is close to 1 (0.93 here). With
slight modifications to the motion, efficiency can be made to approach 1. 
Efficiency increases as the deformation wavelength decreases, so that zero net vertical force and torque are obtained with a purely horizontal motion,
decreasing wasted vertical motion that is not in the direction of locomotion. Efficiency also
increases as the deformation wave is made steeper (body tangent angle approaches $\pm \pi/2$), so transverse motion is aligned with the direction of locomotion. In this limit the body motion is purely
transverse and purely in the direction of motion. Since $\mu_t = 1$, the work done per unit distance
traveled tends to 1.

Figure \ref{fig:OptimalAnisotropicMotionsFig}B shows snapshots when the anisotropy is reversed
($\mu_f = 1 \ll 100 = \mu_t$), so friction is much smaller in the tangential direction (similar to snake robots with wheels whose axes are transverse to the body axis \cite{hopkins2009survey}). Here $\mu_b = 1$ but is arbitrary since there is no backward motion. The body deforms as a sinusoidal leftward moving wave ($\Theta(s,t) = \sin(2\pi(s+t))$). The 
efficiency $\lambda$ is 0.76, and can be made to approach 1 in the limit $\mu_t \to \infty$ by decreasing the amplitude and
the deformation wavelength, so motion is almost purely in the tangential direction and in
the direction of motion. Since $\mu_f = 1$, the work done per unit distance
traveled tends to 1. Unlike in panel A, here the wave shape (whether sinusoidal, triangular, etc.) does
not matter in the limiting case of optimal efficiency. The motions in \ref{fig:OptimalAnisotropicMotionsFig}A and B are
somewhat idealized versions of the direct and retrograde waves shown in figure \ref{fig:PhasePlaneMotionsIsotropicFig} and
are discussed in \cite{AlbenSnake2013}. 
With large backward friction, and $\mu_f \approx \mu_t \approx 1$ ratcheting motions were found to be locally optimal in that work. Now we show that with isotropic friction, none of these motions
is effective. 

\begin{figure} [h]
           \begin{center}
           \begin{tabular}{c}
               \includegraphics[width=6.5in]{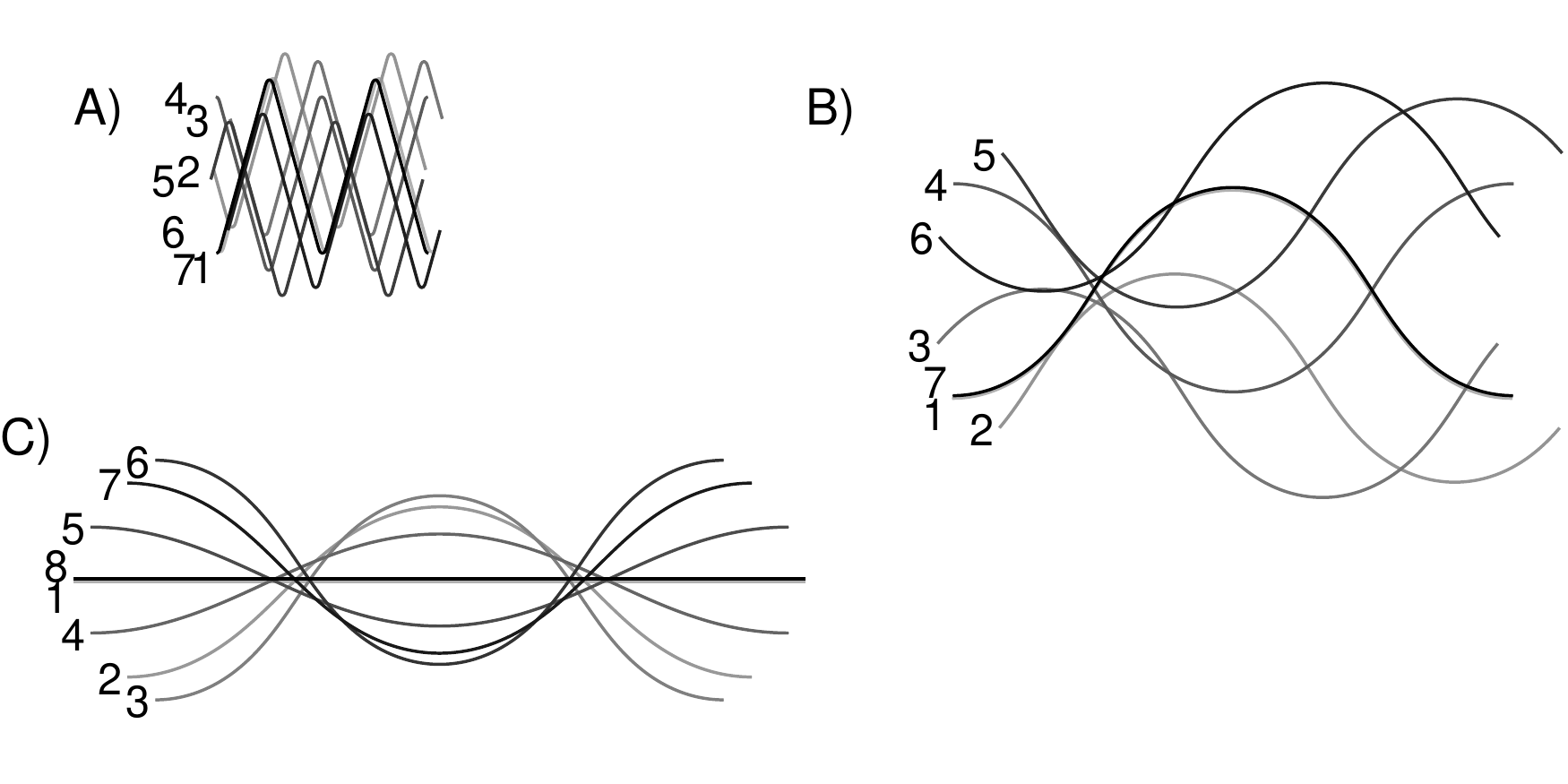} \\
           \vspace{-.25in} \hspace{-.25in}
           \end{tabular}
          \caption{\footnotesize Snapshots of the snake body over one time period 
(darker at later times, labeled near the tail in numerical order) when executing time-periodic body deformations with isotropic friction 
($\mu_t$ = $\mu_f$  = $\mu_b$  = 1).
 A) Traveling wave with a smoothed triangular
body shape (same as in figure \ref{fig:OptimalAnisotropicMotionsFig}A).
B) Traveling wave with a sinusoidal tangent angle profile (same as in figure \ref{fig:OptimalAnisotropicMotionsFig}B). 
C) Standing wave with a sinusoidal tangent angle profile, $\Theta(s,t) = \sin(2\pi s)\sin(2\pi t)$.
 \label{fig:IneffectiveIsotropicMotionsFig}}
           \end{center}
         \vspace{-.10in}
        \end{figure}

In figure \ref{fig:IneffectiveIsotropicMotionsFig}A and B we show snapshots from the
same motions as in figure \ref{fig:OptimalAnisotropicMotionsFig}A and B but with
isotropic friction ($\mu_f = \mu_t = \mu_b = 1$). Panel C shows a standing wave
motion ($\Theta(s,t) = \sin(2\pi s)\sin(2\pi t)$), similar to those which were found to be effective
with large backward friction in \cite{AlbenSnake2013}.
In all three cases the work done against friction is 0.4--0.5 but the distance traveled
is less than 0.005, about the level of numerical error.

\section{Three-link time-harmonic motions \label{sec:3linkharm}}

To increase our intuition about locomotion in the isotropic regime, we now study the efficiency of a
broad range of motions. The space of time-periodic motions $\Theta(s,t)$ is infinite-dimensional,
so to make the problem tractable we look at a finite-dimensional subspace involving
three-link bodies. These have been
studied extensively in locomotion problems in the past (in a viscous fluid at zero Reynolds number)
\cite{Pu1977a,BeKoSt2003a,TaHo2007a,AvRa2008a}. The optimally efficient motion found
in \cite{TaHo2007a} was close to a time-harmonic motion, but with dry friction instead of viscous
forces we have no reason to expect a similar result. In previous work we studied
the motions of 2-link bodies with various friction coefficients, and of 3-link bodies with the
anisotropic friction coefficients measured from real snakes \cite{HuSh2012a} and found locally
optimal motions \cite{JiAl2013}. 
Now with an improved model involving static friction and an improved numerical method
we compute the full range of motions of 3-link bodies with isotropic friction, when the joint angles
are time-harmonic functions. 

\begin{figure} [h]
           \begin{center}
           \begin{tabular}{c}
               \includegraphics[width=5in]{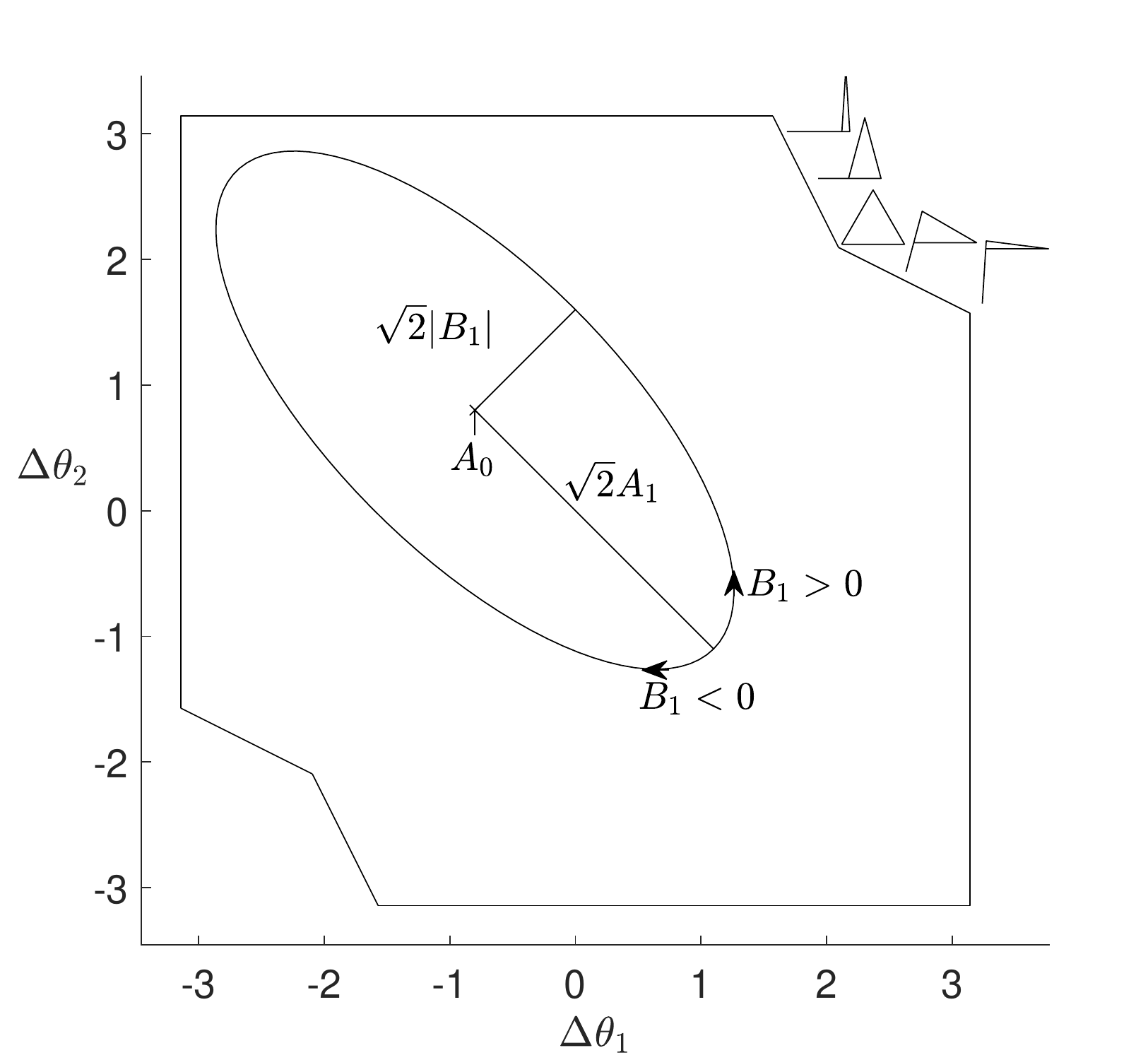} \\
           \vspace{-.25in} \hspace{-.25in}
           \end{tabular}
          \caption{\footnotesize A schematic diagram of an elliptical path in the space of 
non-self-intersecting configurations 
($\Delta\theta_1$, $\Delta\theta_2$) for a three-link body, symmetric about the line
$\Delta\theta_1 = -\Delta\theta_2$. $A_0$ is the average of $\Delta\theta_1$ over
the ellipse and $\sqrt{2}A_1$ and $\sqrt{2}|B_1|$ are the semi-major and
semi-minor axes of the ellipse. The sign of $B_1$ gives the direction in which the
path is traversed. 
 \label{fig:PathsSchematic}}
           \end{center}
         \vspace{-.10in}
        \end{figure}

The bodies' shape at an instant is
described by only two joint angles ($\Delta\theta_1$, $\Delta\theta_2$; see figure
\ref{fig:SmoothAnd3LinkSchematic}) so the possible motions are a set of paths in a 
two-dimensional region shown in figure \ref{fig:PathsSchematic}. The region is
a square with sections removed at the upper right and lower left corners, where
the body self-intersects 
(at the upper right corner, five bodies are shown corresponding to 
configurations along the boundary of this section).  

Within this space of paths, we consider a low-dimensional subspace---motions that have a single frequency 
(i.e. time-harmonic motions)---and are symmetric about
the line $\Delta\theta_1 = -\Delta\theta_2$. This symmetry guarantees no net rotation over a period (see appendix
\ref{norotation}), so
the long-time trajectory of the body is a straight line rather than a circle. 
Such paths are described by 
\begin{align}
\Delta\theta_1(t) = A_0 + A_1\cos(2\pi t) + B_1\sin(2\pi t), \quad
\Delta\theta_2(t) = -A_0 - A_1\cos(2\pi t) + B_1\sin(2\pi t), \quad 0 \leq t \leq 1.
\end{align}
\nn The three parameters $A_0, A_1$, and $B_1$ describe an ellipse with center $(A_0, -A_0)$ and principal
semiaxes $A_0$ and $|B_0|$ (figure \ref{fig:PathsSchematic}). We assume $A_0 \geq 0$ without loss of generality, 
so the motion starts at the lower right region instead of the upper left region of the ellipse 
(but the same path is traversed in either case). The sign of $B_1$ gives the direction (clockwise or
counterclockwise) around the path. Changing the sign of $B_1$ reverses time and thus reverses the
motion (when $\mu_b = \mu_f$, as here), giving the same efficiency.

We compute motions over the region of $(A_0, A_1, B_1)$-space giving admissible paths (ellipses
that lie in the region of figure \ref{fig:PathsSchematic}). To solve a large number of motions quickly,
it is efficient to first compute a velocity map (or ``connection'' \cite{KaMaRoMe2005a,HaCh2010a,JiAl2013})---a map from the shape variables ($\Delta\theta_1$, $\Delta\theta_2$)
and their velocities  ($\dot{\Delta\theta}_1$, $\dot{\Delta\theta}_2$) to the body velocities at the
tail $\{\dot{x}_0(t), \dot{y}_0(t), \dot{\theta}_0(t)\}$, from which we can reconstruct the body motion
via (\ref{dttheta0})-(\ref{dtz0}) at each time and thus the efficiency. Because of the scaling
relation (\ref{scaling}), instead of computing $\{\dot{x}_0(t), \dot{y}_0(t), \dot{\theta}_0(t)\}$
over the four-dimensional space ($\Delta\theta_1$, $\Delta\theta_2$, 
$\dot{\Delta\theta}_1$, $\dot{\Delta\theta}_2$) it is enough to compute the tail velocities over
two three-dimensional spaces ($\Delta\theta_1$, $\Delta\theta_2$, $\dot{\Delta\theta}_1$) with
$|\dot{\Delta\theta}_1| \leq 1$ and $\dot{\Delta\theta}_2 = 1$; 
and ($\Delta\theta_1$, $\Delta\theta_2$, $\dot{\Delta\theta}_2$) with
$\dot{\Delta\theta}_1 = 1$ and $|\dot{\Delta\theta}_2| \leq 1$, and then
obtain the tail velocities at other combinations of 
($\dot{\Delta\theta}_1$, $\dot{\Delta\theta}_2$) by rescaling them into one of these
three-dimensional spaces (if $\mu_b \neq \mu_f$ two additional maps would be needed,
at $\dot{\Delta\theta}_1 = -1$ and $\dot{\Delta\theta}_2 = -1$).

\begin{figure} [h]
           \begin{center}
           \begin{tabular}{c}
               \includegraphics[width=6.5in]{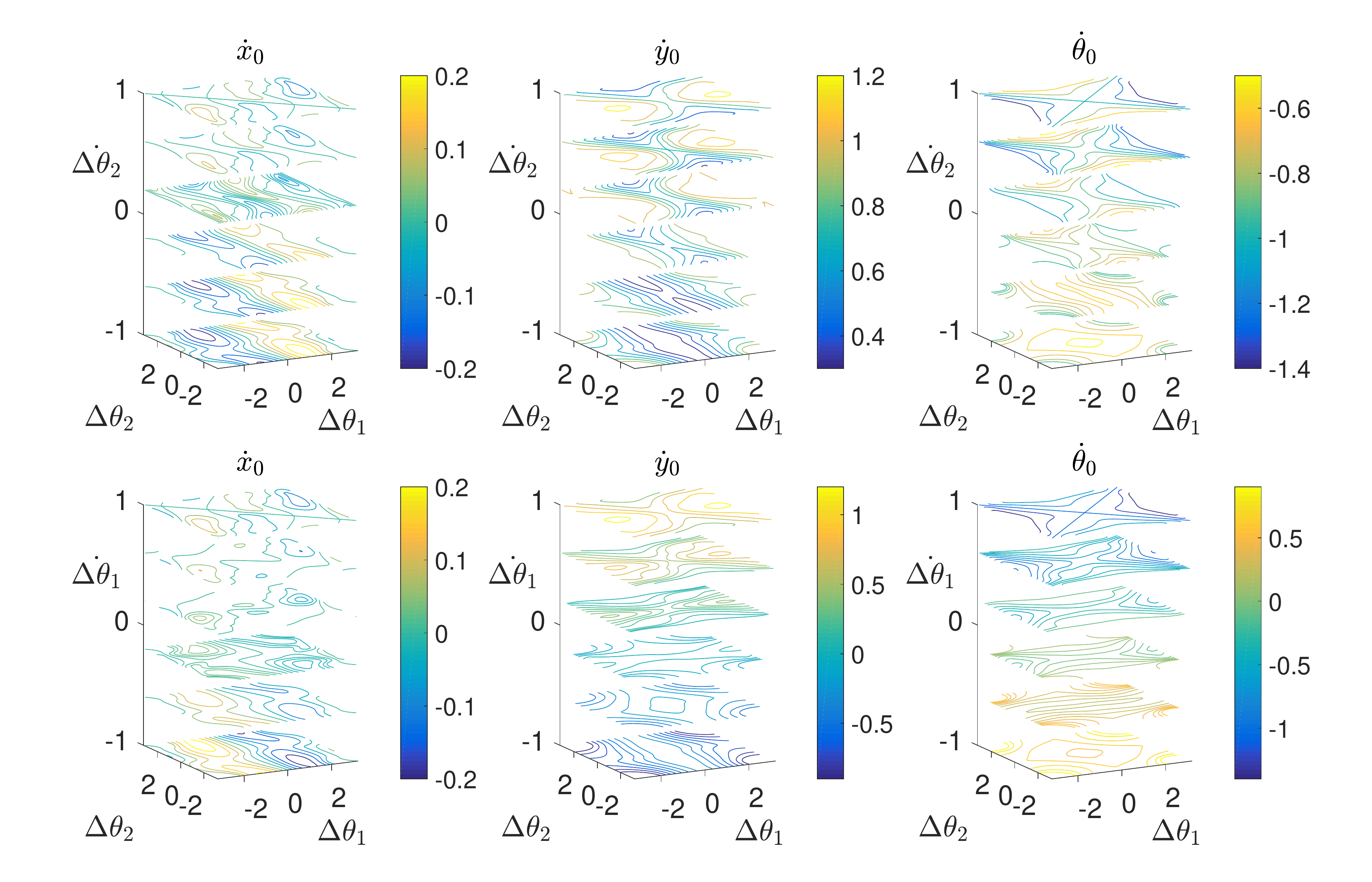} \\
           \vspace{-.25in} \hspace{-.25in}
           \end{tabular}
          \caption{\footnotesize Contour plots showing the three components of the 
velocity map ($\dot{x}_0$, $\dot{y}_0$, and $\dot{\theta}_0$) as functions
of body shape and motion parameters (top row) $\Delta\theta_1$, $\Delta\theta_2$, and $\dot{\Delta\theta_2}$ when 
$\dot{\Delta\theta_1}$ = 1 or (bottom row)  $\Delta\theta_1$, $\Delta\theta_2$, and $\dot{\Delta\theta_1}$ when $\dot{\Delta\theta_2}$ = 1.
 \label{fig:VelocityMapPlots}}
           \end{center}
         \vspace{-.10in}
        \end{figure}
In figure \ref{fig:VelocityMapPlots} we show the two sets of velocity maps used to construct
$\{\dot{x}_0(t), \dot{y}_0(t), \dot{\theta}_0(t)\}$ for any values of 
body shape variables and their velocities when $\dot{\Delta\theta}_1 = 1$ (top row) and
$\dot{\Delta\theta}_2 = 1$ (bottom row). The contours in each slice plane show that the
quantities vary relatively smoothly in these spaces, despite the sharp variations in frictional forces.
We have observed from more extensive data that they are apparently continuous with bounded derivatives,
but that their derivatives change sharply where the regularization parameter is important, i.e. where
static friction plays a role. 

\begin{figure} [h]
           \begin{center}
           \begin{tabular}{c}
               \includegraphics[width=6.5in]{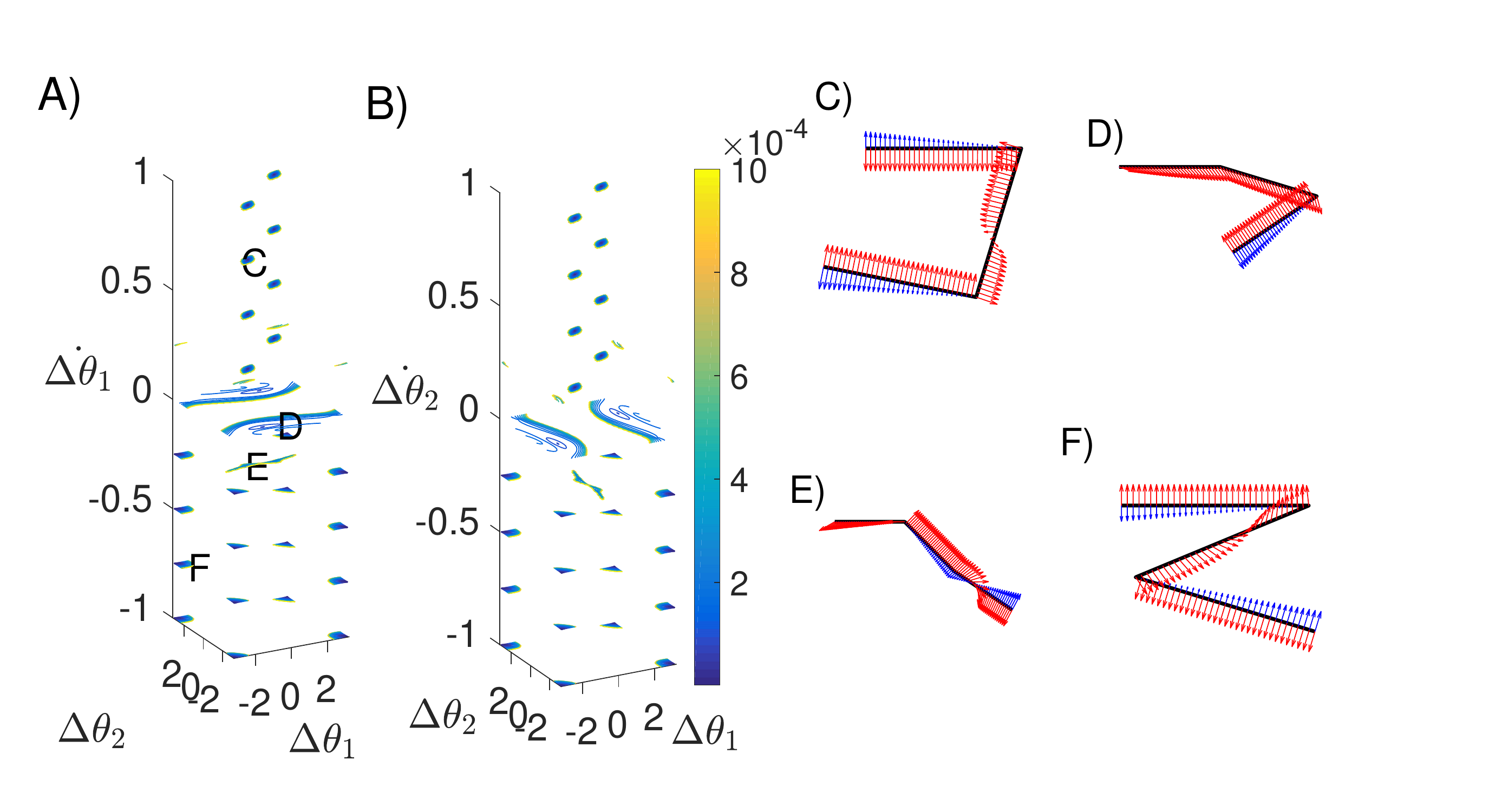} \\
           \vspace{-.25in} \hspace{-.25in}
           \end{tabular}
          \caption{\footnotesize Regions in the space of body shape and motion parameters where
static friction regularization model is involved. A) and B) Contours show regions where the body speed
lies between 0 and $10^{-3}$ over at least one of the three links, in the space of (A) $\Delta\theta_1$, 
$\Delta\theta_2$, and $\dot{\Delta\theta}_2$ when 
$\dot{\Delta\theta}_1$ = 1 or (B)  $\Delta\theta_1$, $\Delta\theta_2$, and $\dot{\Delta\theta}_1$ when $\dot{\Delta\theta}_2$ = 1. C, D, E, and F) Representative examples of body shapes and
motions (labeled in A) where static friction regularization model is involved. Distributions
of body velocities (blue) and frictional forces (red) are shown.
 \label{fig:StaticFrictionPlots}}
           \end{center}
         \vspace{-.10in}
        \end{figure}

Static friction is potentially important when the speed ($\|\partial_t \mathbf{X}\|$) is of the order of the regularization parameter ($\delta = 10^{-4}$) over one or more entire links. If instead small
velocities do not occur, or occur only at discrete points on the body, 
$\delta$ has only a small effect on the net forces and torque.  In figure
\ref{fig:StaticFrictionPlots} we show regions in the velocity map spaces where static friction is important.
Although the regions are small, they are involved in the motions that optimize efficiency, described
in the next section. The regions
can be classified into a small number of cases.  Typical examples are shown in panels C--F, with corresponding
labels in panel A. Case C occurs when $\Delta\theta_1$ and $\Delta\theta_2$ are approximately equal
to $\pi/2$ or $-\pi/2$. The forces from the outer links are nearly equal and opposite, but a small net force
and torque is needed from the middle link to balance those on the outer links. Case D represents a broad region where
one of the link angle velocities ($\dot{\Delta\theta}_1$ or $\dot{\Delta\theta}_2$) is zero and the other
link angle is bent sharply (with magnitude between $\pi/2$ and $\pi$) and has nonzero velocity. Case E represents a smaller
region where one of the link angles has a small but nonzero velocity. Case F occurs when the link angles have magnitudes
near $\pi$ and opposite signs. To understand why static friction is involved we look at cases C and F more closely.

\begin{figure} [h]
           \begin{center}
           \begin{tabular}{c}
               \includegraphics[width=5in]{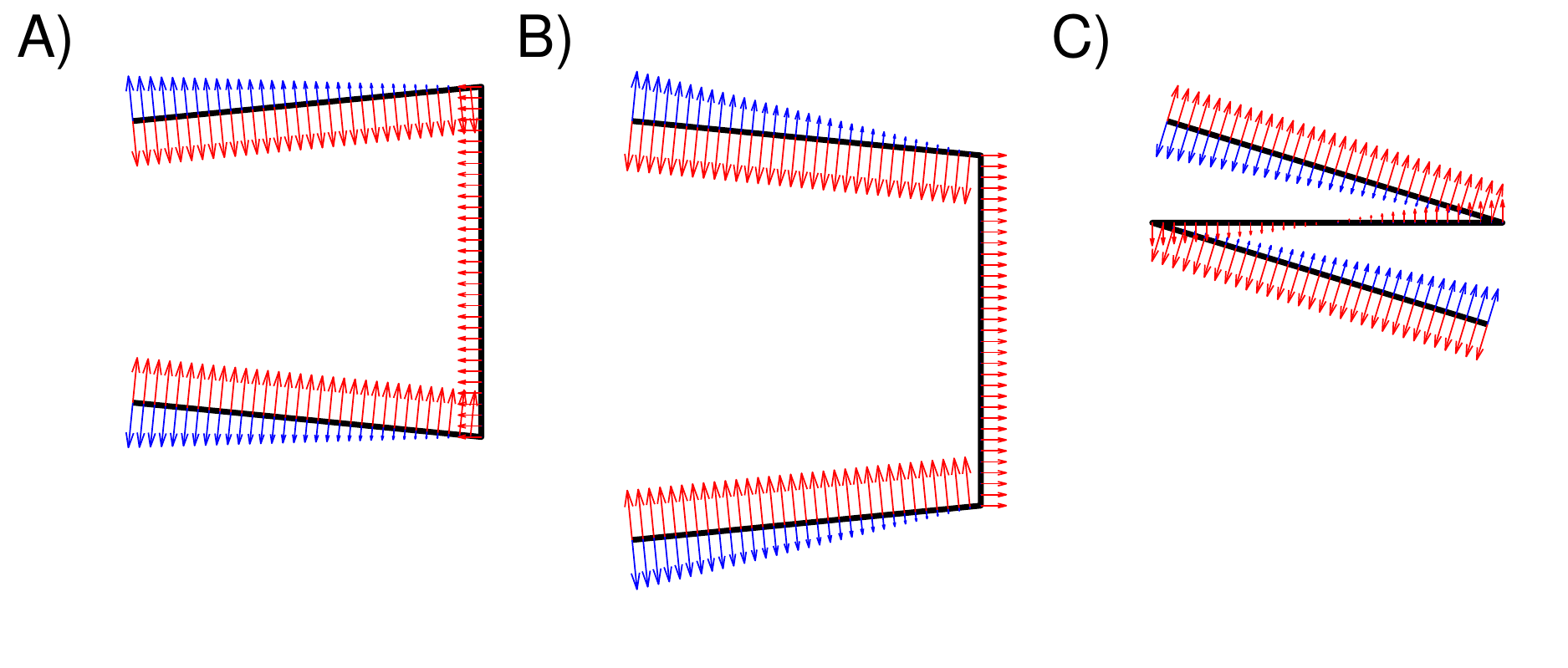} \\
           \vspace{-.25in} \hspace{-.25in}
           \end{tabular}
          \caption{\footnotesize Examples of symmetric body motions where the
static friction regularization model is involved. Distributions
of body velocities (blue) and frictional forces (red) are shown.
 \label{fig:StaticFrictionDiagram}}
           \end{center}
         \vspace{-.10in}
        \end{figure}

Figure \ref{fig:StaticFrictionDiagram}A and B show symmetric examples similar to figures \ref{fig:StaticFrictionPlots}C and \ref{fig:StaticFrictionSketch}.
The outer links provide forces that are nearly opposite and in the vertical direction, but have a small horizontal component.
Due to the top-bottom symmetry of the configuration, the velocity of the middle link can only be horizontal for the vertical
forces to balance. Without regularization, the horizontal force per unit length on the middle link could only be 
0 or $\pm1$, which cannot
balance the small horizontal forces from the outer links. Regularization allows for a smaller horizontal force with a nearly
static middle link, like the force from static friction. Figure \ref{fig:StaticFrictionDiagram}C shows a symmetric version of 
figure \ref{fig:StaticFrictionPlots}F---symmetric with respect to reflection through
the body center. The outer links provide forces that are equal and opposite, but give a small net torque. To provide a torque
with zero net force, the middle link has a purely rotational motion. 
Without regularization the force density on the middle link could only be 0, or -1 on one half and 1 on the other, giving
a net torque of 0 or $\pm1/36$ (since the link has length 1/3). Regularization allows a different torque to be obtained with a nearly
static middle link, like that due to static friction. The other cases in figure \ref{fig:StaticFrictionPlots} are more
difficult to explain because they are not symmetric.

\begin{figure} [h]
           \begin{center}
           \begin{tabular}{c}
               \includegraphics[width=6.5in]{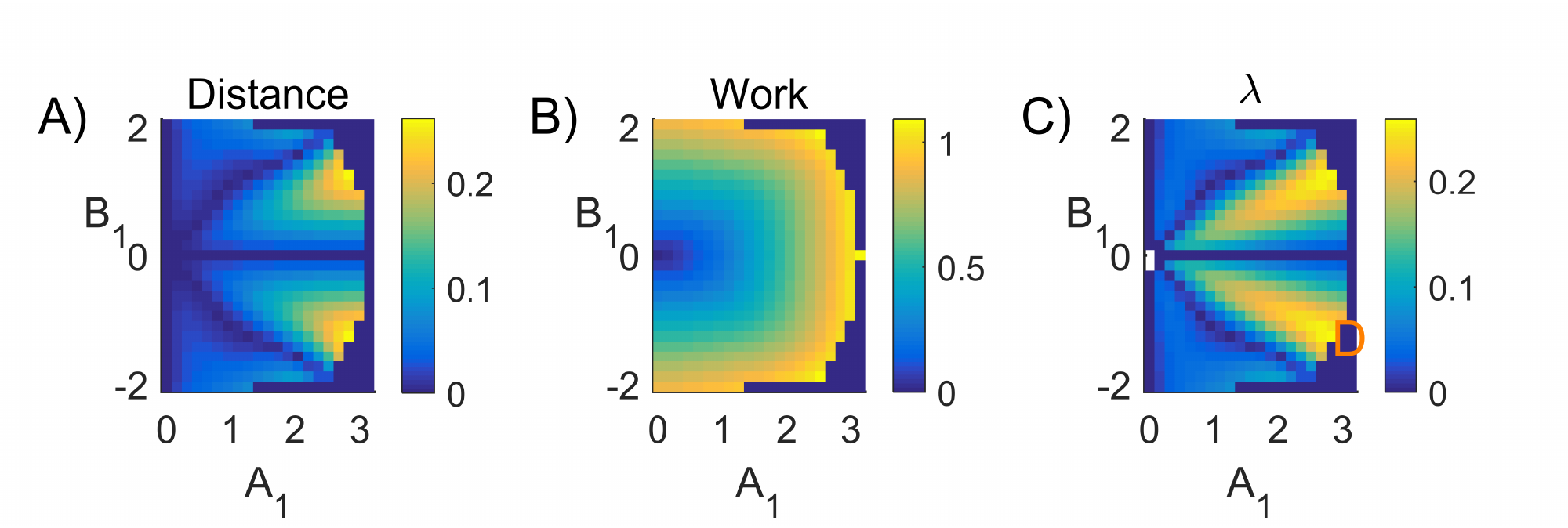} \\
           \vspace{-.25in} \hspace{-.25in}
           \end{tabular}
          \caption{\footnotesize Plots of (A) the distance traveled, (B) the work done against friction, and
(C) the efficiency $\lambda$ (distance/work) for elliptical paths with $A_0 = 0$.
 \label{fig:ZeroA0Fig}}
           \end{center}
         \vspace{-.10in}
        \end{figure}

We now compute the distance traveled, work done, and their ratio $\lambda$, the efficiency, for the elliptical trajectories shown in figure \ref{fig:PathsSchematic}, parametrized by $A_0, A_1,$ and $B_1$. To aid our presentation we begin by showing in figure \ref{fig:ZeroA0Fig} the results in the two-parameter space with $A_0 = 0$. These are for motions that are symmetric with respect to the line
$\Delta\theta_1 = \Delta\theta_2$, but there is no reason {\it a priori} to prefer such motions.

Figure \ref{fig:ZeroA0Fig}A shows that the distance traveled per period is largest for a localized region of 
motions at the limit of self-contact. 
The dark blue region beyond the outer boundary of the shaded region gives coefficients for motions that involve self-contact.
The distance is nearly zero for motions near the line $A_1 = B_1$, i.e. circular trajectories. These trajectories
approximate the traveling-wave motions shown in the previous section, and 
are effective for low Reynolds number swimming \cite{Pu1977a,BeKoSt2003a,TaHo2007a,AvRa2008a} given
the 2:1 drag anisotropy of slender swimming bodies \cite{lauga2009hydrodynamics}. The line
$A_1 = 0$ corresponds to standing wave motions similar to that in the previous section, and results
in zero distance traveled since the motion is the same but the trajectory is reversed under time reversal.
The line $B_1 = 0$ gives standing wave motions that are antisymmetric about the body midpoint but also unchanged under time reversal, and thus also give zero net distance traveled.

Panel B shows the work done
per period, which has a much simpler distribution---it is nearly radially symmetric. Larger coefficients $A_1$ and $B_1$
are clearly correlated with larger sweeping motions of the links. The work done has no obvious relationship with the
distance traveled (A), because the net translation (0.261 body lengths at maximum) 
is only a small contribution to the total motion in most cases. The efficiency (C) has a pattern similar to the distance,
though of course smaller-amplitude motions are weighted more favorably. Nonetheless, the most efficient motion is close to
the distance-maximizing motion, and has efficiency 0.259. The quantities are invariant when the sign of $B_1$ is
changed, because the motion is simply reversed in time.

\begin{figure} [h]
           \begin{center}
           \begin{tabular}{c}
               \includegraphics[width=6.5in]{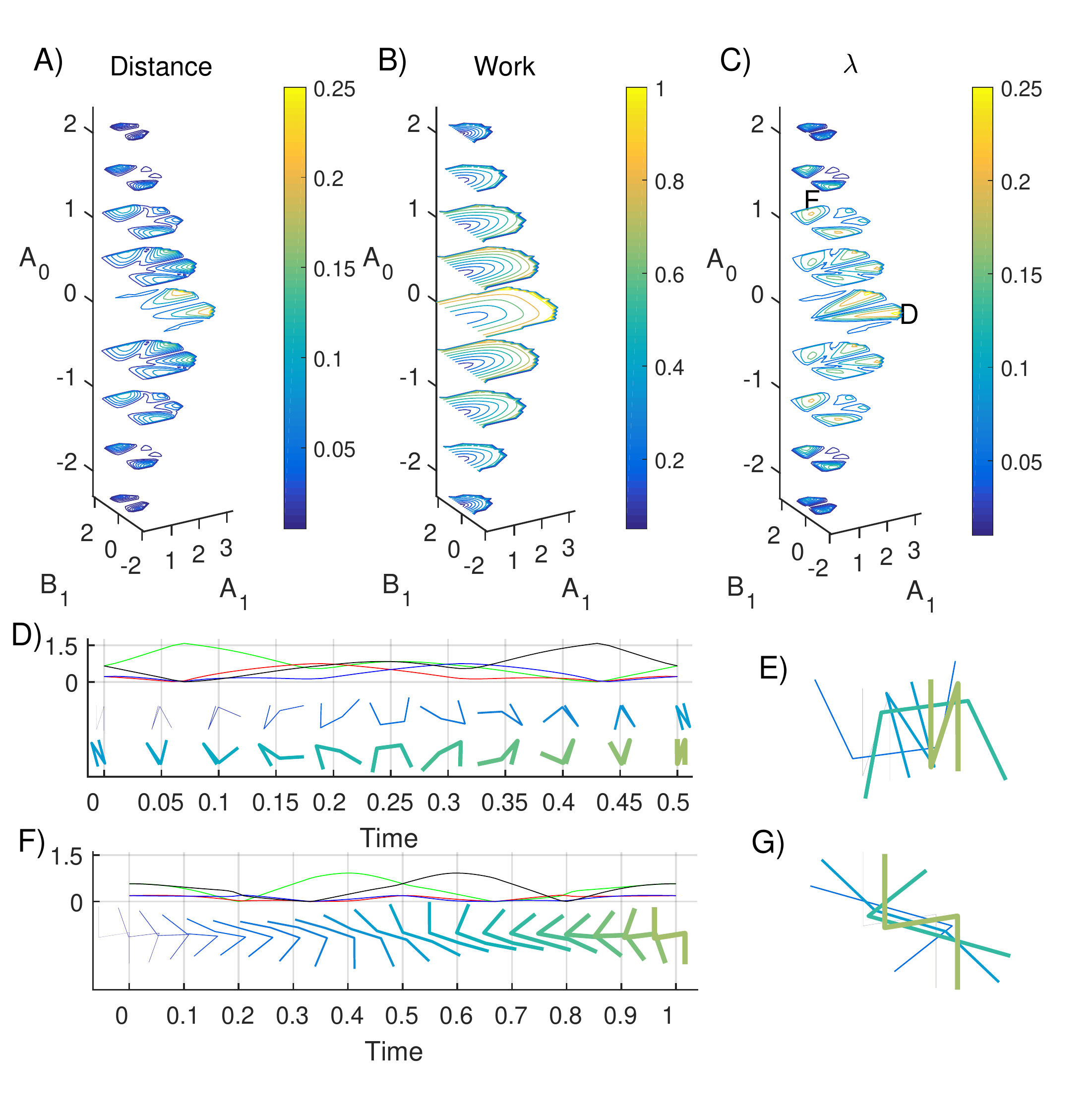} \\
           \vspace{-.25in} \hspace{-.25in}
           \end{tabular}
          \caption{\footnotesize
Contour plots of (A) the distance traveled, (B) the work done against
friction, and
(C) the efficiency $\lambda$ (distance/work) for elliptical paths with various $A_0$.
D) For the most efficient symmetric elliptical body motion (labeled `D' in panel C and
Fig. \ref{fig:ZeroA0Fig}C), the top four lines show the speeds of the four link endpoints (from tail to head: black, blue, red, and
green). Below are snapshots of the body during the first (top
row) and second (bottom row) half-periods. E) A subset of snapshots from panel D in
the lab frame. F)  For the motion giving the 
second best local optimum in efficiency (labeled `F' in panel C),
the same data as in panel D.  G) A subset of snapshots from panel F in
the lab frame.
 \label{fig:Elliptical3LinkMotionsBothSymmFig}}
           \end{center}
         \vspace{-.10in}
        \end{figure}

In figure \ref{fig:Elliptical3LinkMotionsBothSymmFig}A-C we show the same quantities but with
$A_0$ varied over its full range. At the middle of the $A_0$ axis is $A_0 = 0$, so there 
the contour plots show the same data as in the previous figure. When $A_0 = 0$, the
largest distance is achieved at a point with $A_1 > B_1$. As $A_0$ increases or decreases (moving up or down the vertical axis), another local maximum, this one having $B_1 > A_1$ gives a larger distance. In 
panel B, the work maintains an approximate radial symmetry, and does not depend strongly on $A_0$ 
(which varies the offset bias but not the sweeping amplitude of the links' motions). In panel C,
the efficiency has three local maxima. The global maximum is found at $A_0$ = 0, has efficiency
0.259, and is labeled `D' (here and in the previous figure). The motion is shown in panels D and E. The second best local optimum is found at $A_0$ = 1.1, has efficiency 0.207, and is labeled `F'. 
The motion is shown in panels F and G. The third local optimum (not shown) has $A_0$ = 2.5, efficiency 0.094. In panel D, the snapshots of the globally optimal motion are arranged in two rows: first half-period (top) and second half-period (bottom), for which the body shape is a mirror image of that in first half-period ($\Theta$ has opposite sign). The snapshots are shown at equal time intervals during the half-periods (time is labeled at the bottom). At the top are four colored lines showing the speeds of the four endpoints
of the three links versus time for the first half-period. We see that at two times, 0.07 and 0.43, three
of the four endpoints (and two of the three links) are almost static. Here the static friction regularization
is involved in the force balance. At $t$ = 0.07, one link extends rightward while the other two remain
fixed. At $t$ = 0.43, one link is retracted rightward towards the other two. The snapshots
are shown at their true locations in the lab frame in panel E; the body moves about 0.26
body lengths. For the second local optimum, the snapshots are shown in panel F, in time
increments of 0.05 over an entire period. Near $t$ = 0.33 and 0.67, two of the links are almost
static, while the third link moves in the direction of locomotion. The motion is shown in the lab frame in panel G. 
The distance traveled is about 37\% of that in panel E and the work done is about 47\%.
Both of the optimal motions can be described as follows: One of the outer links is rotated forward (i.e. in the direction of
locomotion), with the other two mostly static (for $t$ = 0.38-0.5 in D, 0.2-0.5 in F), then the other outer link is rotated forward with the other two mostly static (from $t$ = 0-0.12 in D, 0.5-0.8 in F), then the middle link is moved, which requires the two outer links to rotate (from $t$ = 0.12-0.38 in D, 0.8-1 and 0-0.2 in F). The motions are roughly speaking similar
to concertina motion, where the snake moves part of its body (like one of the outer links) forward, pushing off
of (or pulling towards) the rest of the body (like the other two links) that is held fixed by static friction, 
forming an ``anchor'' \cite{gray1946mechanism,jayne1986kinematics,jayne1991kinematics}. Because the
body has only three links, moving the middle link forward requires all three links to move and rotate,
so this part of the motion is somewhat distinct.


\section{Optimal motions \label{sec:opt}}

Inspired by the concertina-like motions in the previous section, we now look for more
general smooth motions that can achieve the highest possible efficiency for any inextensible body,
not necessarily one with three links. First, 
we show that an upper bound on efficiency for any motion is the reciprocal of the
smallest friction coefficient (1 in the isotropic case).

The distance traveled by the body (\ref{dist}) is 
the same for all $s$ since the body moves as a translation without rotation after one period.
Thus we can write
\begin{align}
d =\left\Vert\int_0^1 \int_0^1 \partial_t \mathbf{X}(s,t) ds dt\right\Vert.
\end{align}
\nn The work done against friction is (\ref{W})
with $\mathbf{f}$ from (\ref{friction}). Let $u_s \equiv \partial_t \mathbf{X}\cdot \hat{\mathbf{s}}$
and $u_n \equiv \partial_t \mathbf{X}\cdot \hat{\mathbf{n}}$.  We have
\begin{align}
-\mathbf{f}(s,t) \cdot \partial_t \mathbf{X}(s,t) = \frac{\mu_t u_n^2 + \mu_s u_s^2}{\sqrt{u_s^2 + u_n^2}}
\end{align}
\nn where
\begin{align}
\mu_s(s,t) \equiv \left( \mu_f H(\widehat{\partial_t{\mathbf{X}}}\cdot \hat{\mathbf{s}})
+ \mu_b (1-H(\widehat{\partial_t{\mathbf{X}}}\cdot \hat{\mathbf{s}}))\right).
\end{align}
\nn Therefore 
\begin{align}
-\mathbf{f}(s,t) \cdot \partial_t \mathbf{X}(s,t) \geq \min(\mu_t, \mu_f, \mu_b) \sqrt{u_s^2 + u_n^2}
= \min(\mu_t, \mu_f, \mu_b) \|\partial_t \mathbf{X}(s,t)\|
\end{align}
\nn and
\begin{align}
W \geq \min(\mu_t, \mu_f, \mu_b) \int_0^1 \int_0^1 \|\partial_t \mathbf{X}(s,t)\| ds dt
\geq d\,\min(\mu_t, \mu_f, \mu_b).
\end{align}
\nn so
\begin{align}
\lambda =\frac{d}{W} \leq \frac{1}{\min(\mu_t, \mu_f, \mu_b)}. \label{lambdaupper}
\end{align}

This upper bound corresponds to a body that translates uniformly in the direction of lowest friction.
Such a motion cannot have zero net force for nonzero friction, but
we now show simple motions that satisfy the equations of motion and saturate this upper bound in
the limit of a small parameter, for
any choice of friction coefficients, including the isotropic case. These
are concertina-like motions, in the sense that part of the body forms an anchor, remaining static
due to static friction, allowing the rest of the body to be pushed or pulled forward. 

\begin{figure} [h]
           \begin{center}
           \begin{tabular}{c}
               \includegraphics[width=7in]{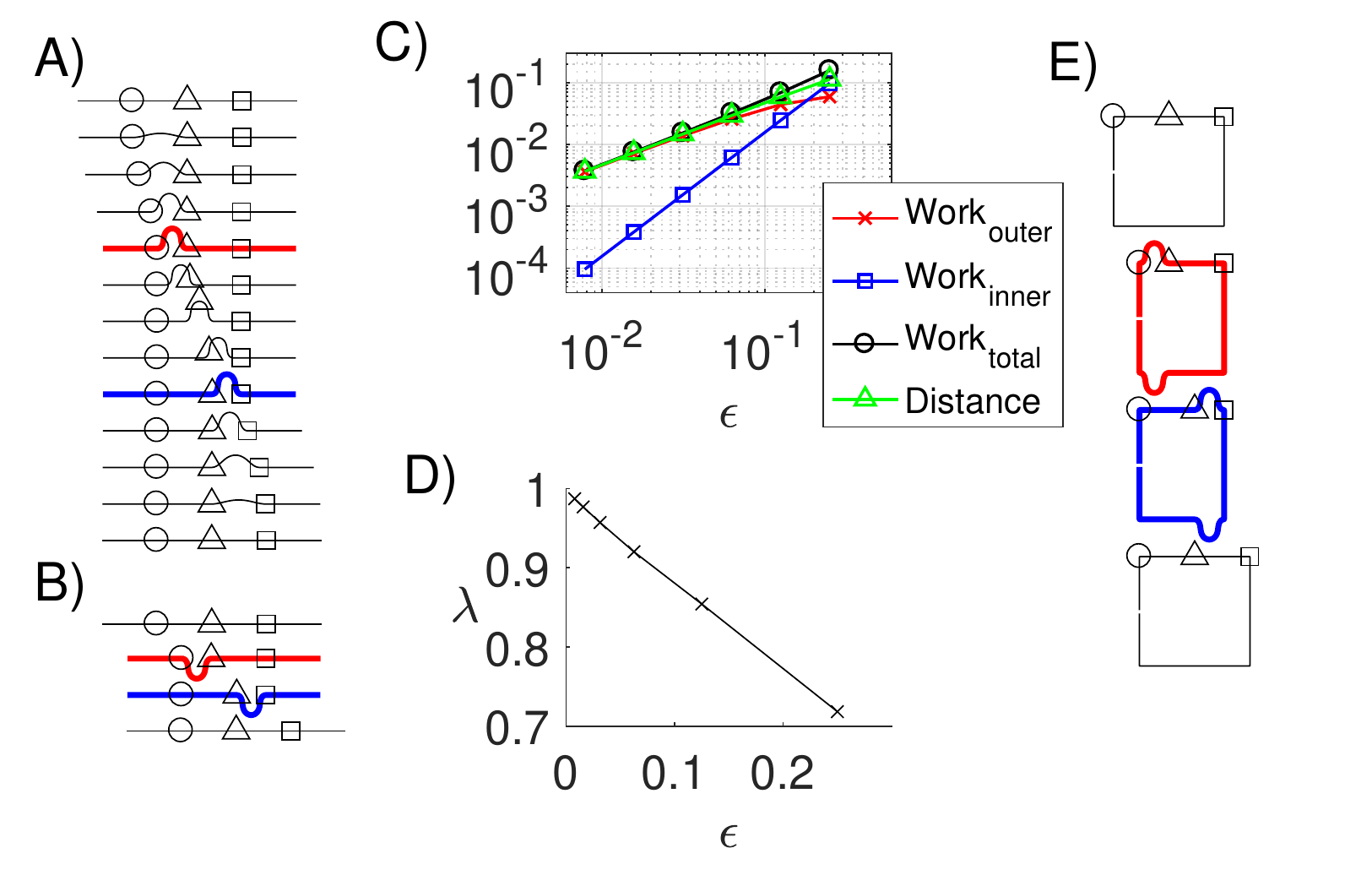} \\
           \vspace{-.25in} \hspace{-1.5in}
           \end{tabular}
          \caption{\footnotesize Motions and performance of optimally efficient crawlers. A) Snapshots
of the crawler during the first half-period of motion. The circle and square are the endpoints of the bump
region, and the triangle is the body midpoint. B) Snapshots during the second half period.
C) Plot of work done against friction 
over one period outside the bump region (``Work\textsubscript{outer}", red), that done
inside the bump region (``Work\textsubscript{inner}", blue), the total work 
 (black), and the distance covered in one period (green), versus the bump region width
parameter $\epsilon$. D) The efficiency $\lambda =$ Work/Distance versus the
bump region length $\epsilon$.
E) Snapshots of an efficient (symmetric) crawler when $\mu_t$ is the smallest friction coefficient.
 \label{fig:OptimalCrawlersFig}}
           \end{center}
         \vspace{-.10in}
        \end{figure}

We first assume isotropic friction. The body is initially straight (see figure \ref{fig:OptimalCrawlersFig}A, top). 
The motion has three stages. In stage one, a straight segment in the rear half of the body but near the midpoint (between
the circle and triangle in figure \ref{fig:OptimalCrawlersFig}A) forms a ``bump.'' It deforms from straight to curved, but keeping the tangent angles at its endpoints unchanged, so the endpoints get closer. This pulls the rear of the body forward, because the front portion (front half) of the body (the ``anchor") is static due to
static friction. If the front portion of the body slides with an $O(1)$ velocity, the rear portion of the body is not large
enough to provide a balancing force. Therefore, the front portion of the body's velocity is O($\delta$). At the
end of stage one (red body in panel A), the bump reaches its maximum amplitude. In stage two (from the red body to the blue body), the bump travels forward along the body, to the region between the triangle and the square. The blue shape
is thus a mirror image of the red shape. Here the body endpoints do not move, because the region away from the
bump (left of the circle and right of the square) is an anchor. Stage three 
(from the blue body to the last straight configuration in A) is essentially the reverse of stage one---the bump flattens out,
pushing the region in front of the square forward, with the back region of the body fixed because now it is an anchor.
The net result is that the body has moved rightward some amount (which can be seen comparing the body endpoints
over the sequence of motions). In addition to moving rightward, the body undergoes a much smaller vertical
displacement and rotation because the bump is upward. To achieve a motion with zero net rotation (and zero net vertical
displacement), we then perform the mirror image of the motion (panel B) for $0.5 \leq t \leq 1$, with $\Theta(s,t+0.5)=-\Theta(s,t)$.
Then we see that the mirror image motion in the lab frame is a solution:
\begin{align} 
\dot{\theta}_0(t+0.5) &= -\dot{\theta}_0(t), \partial_t \theta(s,t+0.5) =-\partial_t \theta(s,t),  \\
\dot{y}_0(t+0.5) &= -\dot{y}_0(t), \partial_t y(s,t+0.5) = -\partial_t y(s,t), \\
\dot{x}_0(t+0.5) &= \dot{x}_0(t), \partial_t x(s,t+0.5) = \partial_t x(s,t).
\end{align} 
\nn We have the same horizontal displacement but the vertical displacement and rotation are reversed. Panel B
shows the snapshots in the simulation of the second half of the motion (at the beginning/end of the
three stages only).
The length of the bump (half the arc-length distance from the circle to the square) 
is a control parameter $\epsilon$
that we can shrink to zero. We show now that the distance traveled is proportional to $\epsilon$,
and the work done can be decomposed into two parts. The work done inside the bump region (left of the circle 
and right of the square) is proportional to $\epsilon^2$ (blue squares in panel C). The
velocities in the bump region $\sim \epsilon$, the frictional force density $\sim 1$, 
and the bump region length $\sim \epsilon$,
so by (\ref{W})
\begin{align}
W_{\mbox{inner}} \sim  \int_0^1 
\int_{0.5-\epsilon}^{0.5+\epsilon} 1 \cdot \epsilon ds dt = O(\epsilon^2). \label{Winner}
\end{align}
\nn The work done outside the bump region ($W_{\mbox{outer}}$, red crosses in C) approaches the distance traveled (green triangles) as $\epsilon \to 0$, and both
are proportional to $\epsilon$. $W_{\mbox{outer}}$ is approximately the unit frictional
force density times the body speed 
in the region outside the bump multiplied by the length of that
region $\sim 1$:
\begin{align}
W_{\mbox{outer}} \sim  \int_0^1 
\int_{\{0 \leq s \leq 0.5-\epsilon\} \cup \{0.5+\epsilon \leq s \leq 1\}} 1 \cdot \partial_t x(s,t) ds dt \sim d. \label{Wouter}
\end{align}
\nn Adding (\ref{Winner}) and (\ref{Wouter}) we have $\lambda = 1 + O(\epsilon)$. This is shown in panel D for the motions in panels A-B. When $\epsilon$ decreases
below 0.1, we find it is necessary to decrease the numerical regularization parameter $\delta$ from $10^{-4}$ to
$10^{-6}$ or $10^{-8}$ so it does not affect the results (i.e. so $\delta$ is much smaller than the typical
speed of body deformation).

Now assume the friction coefficients are anisotropic. 
If the smallest friction coefficient is either $\mu_f$ or $\mu_b$, then the body should be oriented in panels A-B so that
the lower of $\mu_f$ and $\mu_b$ applies for motion to the right. If instead the smallest friction coefficient is $\mu_t$, then
we bend the body so that it has two bump regions, and the outer regions are oriented transverse to the direction of locomotion (see
figure \ref{fig:OptimalCrawlersFig}E).
By symmetry, motion is solely in the horizontal direction (the mirror image stroke in panel B is not required now). 
Snapshots are shown only at the beginning/end of each stage in panel E. With anisotropic friction, 
the above
estimate for $W_{\mbox{inner}}$ (\ref{Winner}) is multiplied by $\max(\mu_t, \mu_f, \mu_b)$ to obtain
an upper bound, while that for $W_{\mbox{outer}}$ (\ref{Wouter}) is 
multiplied by $\min(\mu_t, \mu_f, \mu_b)$. The global upper bound
for $\lambda$ (\ref{lambdaupper}) is achieved in the limit $\epsilon \to 0$.

We have assumed an inextensible body. For an extensible body, a one-dimensional version of the above motion is 
obtained
by projecting the body density distribution at each instant onto the horizontal axis.  Similar longitudinal motions are
used by certain soft-bodied animals (e.g. worms) that alternately contract and extend longitudinal muscles 
\cite{keller1983crawling}. Snakes, however, are nearly inextensible due to their backbone \cite{lillywhite2014snakes}.

\section{Conclusion}

In this work we have studied the locomotion of bending and sliding bodies under isotropic friction.
We developed a regularization approach to handle cases where static friction is needed to find a solution. We also introduced a
fixed-point iteration method that can compute the body tail velocities robustly from all initial guesses 
without the need for a line search method. 
We first used the method to show that the most efficient motions with anisotropic friction---traveling wave deformations---lead to little or no 
locomotion with isotropic friction.
Next, we used the method to compute the velocity map for the three-dimensional
body shape and shape velocity spaces of a three-link crawler. We used these maps to obtain a general picture of the
locomotion efficiency landscape for the 3D space of coefficients giving symmetrical elliptical paths in the space of the 
body link angles. We found that static friction regularization is involved in small (but important) regions of the
velocity map and described their necessity in symmetric cases.
The distance traveled and efficiency are very small for motions corresponding to standing waves or
traveling waves. The efficiency has three local maxima, and the top two (0.21 and 0.26) occur at 
motions that are similar to concertina locomotion---a sequence of
motions in which one of the links moves forward while the other two links remain almost motionless.

We then proposed a class of concertina-like motions that saturate the upper bound for efficiency for any
choice of friction coefficients.  The
optimal smooth motions of section \ref{sec:opt} require short wavelengths $\sim \epsilon$ (and large
frequencies $\sim 1/\epsilon$ to travel an $O(1)$ distance), which explains why 
the numerical optimization
using 45 or 190 modes in \cite{AlbenSnake2013} did not converge to such motions.
It is interesting, however, 
that in the optimal time-harmonic motions with only three links, concertina-like motions can be seen. 
Although static friction arises in the optimal motions shown here, we believe that solutions
with similar motions---and similar efficiencies---may exist with only the kinetic friction model (i.e. without
regularization). In other words, the motion may be altered so that instead of remaining static,
the ``anchor" portion of the body slides slowly but has enough kinetic friction to balance that on
the remainder of the body.





\begin{acknowledgments}
This research was supported by the NSF Mathematical Biology program under
award number DMS-1811889.
\end{acknowledgments}

\appendix
\section{Zero net rotation for motions symmetric with respect to $\Delta\theta_1 = -\Delta\theta_2$.
\label{norotation}}

We show here that motions of three-link bodies that are symmetric with respect to the line
$\Delta\theta_1 = -\Delta\theta_2$ (e.g. figure \ref{fig:PathsSchematic}) result in zero net rotation over a period. 
For such motions we can assume (as in section \ref{sec:3linkharm}) that the body motion starts on the line $\Delta\theta_1 = 
-\Delta\theta_2$ in configuration space (by shifting time by a constant if necessary), so the body lies on this line at $t$ = 0 and 1, and at $t$ = 1/2 by the symmetry of the path.
The symmetry implies
that the link angle differences at $t$ and $1-t$ are related by $\Delta\theta_1(t) = - \Delta\theta_2(1-t)$
and $\Delta\theta_2(t) = - \Delta\theta_1(1-t)$. Thus if the three links at time $t$ have tangent angles 
$\{ \theta_0(t), \theta_0(t) + \Delta\theta_1(t), \theta_0(t) + \Delta\theta_1(t) + \Delta\theta_2(t)\}$ 
then those at $1-t$ have tangent angles
$\{ \theta_0(1-t), \theta_0(1-t) - \Delta\theta_2(t), \theta_0(1-t) - \Delta\theta_2(t) - \Delta\theta_1(t)\}$.
This implies that $\theta(s,t) - \theta(1-s,1-t) = \theta_0(t) + \Delta\theta_1(t) + \Delta\theta_2(t) -\theta_0(1-t)$,
which is independent of $s$. In other words, the body at time $1-t$ has the same shape (tangent angle) as that
at time $t$, up to an overall rotation, when the body at time $1-t$ is viewed from the opposite end---starting 
at $s = 1$ and ending at $s = 0$. If we define a new coordinate $u = 1-s$, we can describe the tangent angle
at time $1-t$ in a body frame running from $u =0$ to $u = 1$ using the function $\Theta_u(u,t)$ as
\begin{align}
\theta(u,1-t) = \theta_{u = 0}(1-t) + \Theta_u(u,1-t).
\end{align}
\nn We have  $\Theta(s,t) = \Theta_u(1-s,1-t)$ and $\partial_t \Theta(s,t) = -\partial_t \Theta_u(1-s,1-t)$,
so in the body frames the two shapes are the same and their rates of change are opposite. Therefore, following
the solution procedure described below equations (\ref{dttheta0})-(\ref{dtz0}),
the solutions for the rotation rates at $s=0$ and $u = 0$ are opposite (if $\mu_b = \mu_f$):
\begin{align}
\dot{\theta}_{s=0,b}(t) = -\dot{\theta}_{u=0,b}(1-t)
\end{align}
\nn Here $b$ denotes body frame, but these are also the rotation rates in the lab frame as discussed
below equations (\ref{dttheta0})-(\ref{dtz0}):
\begin{align}
\dot{\theta}_{s=0}(t) = \dot{\theta}_{s=0,b}(t) = -\dot{\theta}_{u=0,b}(1-t) = -\dot{\theta}_{u=0}(1-t).
\end{align}
\nn We can use these results to compute the net rotation from $t = 0$ to $t = 1$ (over a period),
$\theta_{s=0}(1) - \theta_{s=0}(0)$.
Since the body has $\Delta\theta_1 = -\Delta\theta_2$ at $t$ = 0, 1/2, and 1, 
at those times the tangent angle at $u = 0$ (in the direction of
increasing $u$) is that at $s =  0$ plus $\pi$:
\begin{align}
\theta_{s=0}(1) - \theta_{s=0}(0) &=  \theta_{u=0}(1) + \pi - \theta_{u=0}(1/2) + \theta_{u=0}(1/2) - \theta_{s=0}(0) \\
&= \theta_{u=0}(1) - \theta_{u=0}(1/2) + \theta_{s=0}(1/2) - \theta_{s=0}(0) \\
&= \int_{1/2}^1 \dot{\theta}_{u=0}(t) dt + \int_{0}^{1/2} \dot{\theta}_{s=0}(t) dt \\
&= \int_{1/2}^1 -\dot{\theta}_{s=0}(1-t) dt + \int_{0}^{1/2} \dot{\theta}_{s=0}(t) dt \\
&= \int_{1/2}^0 \dot{\theta}_{s=0}(w) dw + \int_{0}^{1/2} \dot{\theta}_{s=0}(t) dt \\
& = 0,
\end{align}
\nn where $w = 1-t$. 
In words, whatever rotation occurs from $t = 0$ to 1/2 is undone from 1/2 to 1, when we view
the body from the opposite end.

\bibliographystyle{unsrt}
\bibliography{snake}

\begin{thebibliography}{10}

\bibitem{gray1946mechanism}
J~Gray.
\newblock {The mechanism of locomotion in snakes}.
\newblock {\em Journal of Experimental Biology}, 23(2):101--120, 1946.

\bibitem{gray1950kinetics}
J~Gray and HW~Lissmann.
\newblock The kinetics of locomotion of the grass-snake.
\newblock {\em Journal of Experimental Biology}, 26(4):354--367, 1950.

\bibitem{jayne1986kinematics}
Bruce~C Jayne.
\newblock {Kinematics of terrestrial snake locomotion}.
\newblock {\em Copeia}, pages 915--927, 1986.

\bibitem{socha2002kinematics}
John~J Socha.
\newblock {Kinematics: Gliding flight in the paradise tree snake}.
\newblock {\em Nature}, 418(6898):603--604, 2002.

\bibitem{HaCh2010b}
R~L Hatton and H~Choset.
\newblock {Generating gaits for snake robots: annealed chain fitting and
  keyframe wave extraction}.
\newblock {\em Autonomous Robots}, 28(3):271--281, 2010.

\bibitem{MaHu2012a}
Hamidreza Marvi and David~L Hu.
\newblock Friction enhancement in concertina locomotion of snakes.
\newblock {\em Journal of The Royal Society Interface}, 9(76):3067--3080, 2012.

\bibitem{dickinson2000animals}
Michael~H Dickinson, Claire~T Farley, Robert~J Full, MAR Koehl, Rodger Kram,
  and Steven Lehman.
\newblock {How animals move: an integrative view}.
\newblock {\em Science}, 288(5463):100--106, 2000.

\bibitem{hohenegger2010stability}
Christel Hohenegger and Michael~J Shelley.
\newblock Stability of active suspensions.
\newblock {\em Physical Review E}, 81(4):046311, 2010.

\bibitem{olson2011coupling}
Sarah~D Olson, Susan~S Suarez, and Lisa~J Fauci.
\newblock Coupling biochemistry and hydrodynamics captures hyperactivated sperm
  motility in a simple flagellar model.
\newblock {\em Journal of theoretical biology}, 283(1):203--216, 2011.

\bibitem{lim2012fluid}
Sookkyung Lim and Charles~S Peskin.
\newblock Fluid-mechanical interaction of flexible bacterial flagella by the
  immersed boundary method.
\newblock {\em Physical Review E}, 85(3):036307, 2012.

\bibitem{jones2016bristles}
Shannon~K Jones, Young~JJ Yun, Tyson~L Hedrick, Boyce~E Griffith, and Laura~A
  Miller.
\newblock Bristles reduce the force required to ‘fling’wings apart in the
  smallest insects.
\newblock {\em Journal of Experimental Biology}, 219(23):3759--3772, 2016.

\bibitem{bar2005biomimetics}
Yoseph Bar-Cohen.
\newblock {\em Biomimetics: biologically inspired technologies}.
\newblock CRC Press, 2005.

\bibitem{jakimovski2011biologically}
Bojan Jakimovski.
\newblock {\em Biologically inspired approaches for locomotion, anomaly
  detection and reconfiguration for walking robots}, volume~14.
\newblock Springer, 2011.

\bibitem{roper2011review}
DT~Roper, S~Sharma, R~Sutton, and P~Culverhouse.
\newblock A review of developments towards biologically inspired propulsion
  systems for autonomous underwater vehicles.
\newblock {\em Proceedings of the Institution of Mechanical Engineers, Part M:
  Journal of Engineering for the Maritime Environment}, 225(2):77--96, 2011.

\bibitem{jacob1977evolution}
Francois Jacob.
\newblock Evolution and tinkering.
\newblock {\em Science}, 196(4295):1161--1166, 1977.

\bibitem{alexander1996optima}
R~McNeill Alexander.
\newblock {\em Optima for animals}.
\newblock Princeton University Press, 1996.

\bibitem{langerhans2010ecology}
R~Brian Langerhans and David~N Reznick.
\newblock Ecology and evolution of swimming performance in fishes: predicting
  evolution with biomechanics.
\newblock {\em Fish locomotion: an eco-ethological perspective}, pages
  200--248, 2010.

\bibitem{BeKoSt2003a}
L~E Becker, S~A Koehler, and H~A Stone.
\newblock {On self-propulsion of micro-machines at low Reynolds number:
  Purcell's three-link swimmer}.
\newblock {\em Journal of Fluid Mechanics}, 490(1):15--35, 2003.

\bibitem{AvGaKe2004a}
J~E Avron, O~Gat, and O~Kenneth.
\newblock {Optimal swimming at low Reynolds numbers}.
\newblock {\em Physical Review Letters}, 93(18):186001, 2004.

\bibitem{TaHo2007a}
D~Tam and A~E Hosoi.
\newblock {Optimal stroke patterns for Purcell's three-link swimmer}.
\newblock {\em Physical Review Letters}, 98(6):68105, 2007.

\bibitem{fu2007theory}
Henry~C Fu, Thomas~R Powers, and Charles~W Wolgemuth.
\newblock {Theory of swimming filaments in viscoelastic media}.
\newblock {\em Physical Review Letters}, 99(25):258101, 2007.

\bibitem{spagnolie2010optimal}
Saverio~E Spagnolie and Eric Lauga.
\newblock {The optimal elastic flagellum}.
\newblock {\em Physics of Fluids}, 22:031901, 2010.

\bibitem{crowdy2011two}
Darren Crowdy, Sungyon Lee, Ophir Samson, Eric Lauga, and AE~Hosoi.
\newblock {A two-dimensional model of low-Reynolds number swimming beneath a
  free surface}.
\newblock {\em Journal of Fluid Mechanics}, 681:24--47, 2011.

\bibitem{bittner2018geometrically}
Brian Bittner, Ross~L Hatton, and Shai Revzen.
\newblock Geometrically optimal gaits: a data-driven approach.
\newblock {\em Nonlinear Dynamics}, 94(3):1933--1948, 2018.

\bibitem{lighthill1975mathematica}
James Lighthill.
\newblock {\em {Mathematical Biofluiddynamics}}.
\newblock SIAM, 1975.

\bibitem{childress1981mechanics}
Stephen Childress.
\newblock {\em {Mechanics of swimming and flying}}.
\newblock Cambridge University Press, 1981.

\bibitem{sparenberg1994hydrodynamic}
JA~Sparenberg.
\newblock {\em {Hydrodynamic Propulsion and Its Optimization:(Analytic
  Theory)}}, volume~27.
\newblock Kluwer Academic Pub, 1994.

\bibitem{alben2009passive}
S.~Alben.
\newblock {{Passive and active bodies in vortex-street wakes}}.
\newblock {\em Journal of Fluid Mechanics}, 642:95--125, 2009.

\bibitem{michelin2009resonance}
S{\'e}bastien Michelin and Stefan G~Llewellyn Smith.
\newblock {Resonance and propulsion performance of a heaving flexible wing}.
\newblock {\em Physics of Fluids}, 21:071902, 2009.

\bibitem{peng2012bb}
J.~Peng and S.~Alben.
\newblock {Effects of shape and stroke parameters on the propulsion performance
  of an axisymmetric swimmer}.
\newblock {\em Bioinspiration and Biomimetics}, 7:016012, 2012.

\bibitem{gazzola2015gait}
Mattia Gazzola, M{\'e}d{\'e}ric Argentina, and Lakshminarayanan Mahadevan.
\newblock Gait and speed selection in slender inertial swimmers.
\newblock {\em Proceedings of the National Academy of Sciences},
  112(13):3874--3879, 2015.

\bibitem{GuMa2008a}
Z~V Guo and L~Mahadevan.
\newblock {Limbless undulatory propulsion on land}.
\newblock {\em Proceedings of the National Academy of Sciences}, 105(9):3179,
  2008.

\bibitem{aguilar2016review}
Jeffrey Aguilar, Tingnan Zhang, Feifei Qian, Mark Kingsbury, Benjamin McInroe,
  Nicole Mazouchova, Chen Li, Ryan Maladen, Chaohui Gong, Matt Travers, Ross~L
  Hatton, Howie Choset, Paul~B Umbanhowar, and Daniel~I Goldman.
\newblock A review on locomotion robophysics: the study of movement at the
  intersection of robotics, soft matter and dynamical systems.
\newblock {\em Reports on Progress in Physics}, 79(11), 2016.

\bibitem{lillywhite2014snakes}
Harvey~B Lillywhite.
\newblock {\em How Snakes Work: Structure, Function and Behavior of the World's
  Snakes}.
\newblock Oxford University Press, 2014.

\bibitem{jayne1988mechanical}
Bruce~C Jayne.
\newblock Mechanical behaviour of snake skin.
\newblock {\em Journal of Zoology}, 214(1):125--140, 1988.

\bibitem{seigel1987snakes}
Richard~A Seigel, Joseph~T Collins, and Susan~S Novak.
\newblock {\em Snakes: ecology and evolutionary biology}.
\newblock Macmillan New York etc., 1987.

\bibitem{HuNiScSh2009a}
D~L Hu, J~Nirody, T~Scott, and M~J Shelley.
\newblock {The mechanics of slithering locomotion}.
\newblock {\em Proceedings of the National Academy of Sciences}, 106(25):10081,
  2009.

\bibitem{HuSh2012a}
D~L Hu and M~Shelley.
\newblock {Slithering Locomotion}.
\newblock In {\em Natural Locomotion in Fluids and on Surfaces}, pages
  117--135. Springer, 2012.

\bibitem{goldman2010wiggling}
Daniel~I Goldman and David~L Hu.
\newblock Wiggling through the world: The mechanics of slithering locomotion
  depend on the surroundings.
\newblock {\em American Scientist}, 98(4):314--323, 2010.

\bibitem{transeth2008snake}
Aksel~Andreas Transeth, Remco~I Leine, Christoph Glocker, Kristin~Ytterstad
  Pettersen, and P{\aa}l Liljeb{\"a}ck.
\newblock Snake robot obstacle-aided locomotion: Modeling, simulations, and
  experiments.
\newblock {\em IEEE Transactions on Robotics}, 24(1):88--104, 2008.

\bibitem{transeth2009survey}
Aksel~Andreas Transeth, Kristin~Ytterstad Pettersen, and P{\aa}l Liljeb{\"a}ck.
\newblock A survey on snake robot modeling and locomotion.
\newblock {\em Robotica}, 27(07):999--1015, 2009.

\bibitem{ohno2001design}
Hidetaka Ohno and Shigeo Hirose.
\newblock Design of slim slime robot and its gait of locomotion.
\newblock In {\em Intelligent Robots and Systems, 2001. Proceedings. 2001
  IEEE/RSJ International Conference on}, volume~2, pages 707--715. IEEE, 2001.

\bibitem{liljeback2012review}
P{\aa}l Liljeb{\"a}ck, Kristin~Ytterstad Pettersen, {\O}yvind Stavdahl, and
  Jan~Tommy Gravdahl.
\newblock A review on modelling, implementation, and control of snake robots.
\newblock {\em Robotics and Autonomous Systems}, 60(1):29--40, 2012.

\bibitem{chernousko2005modelling}
Felix~L Chernousko.
\newblock Modelling of snake-like locomotion.
\newblock {\em Applied mathematics and computation}, 164(2):415--434, 2005.

\bibitem{wagner2013crawling}
Gregory~L Wagner and Eric Lauga.
\newblock Crawling scallop: Friction-based locomotion with one degree of
  freedom.
\newblock {\em Journal of theoretical biology}, 324:42--51, 2013.

\bibitem{lauga2009hydrodynamics}
Eric Lauga and Thomas~R Powers.
\newblock The hydrodynamics of swimming microorganisms.
\newblock {\em Reports on Progress in Physics}, 72(9):096601, 2009.

\bibitem{AlbenSnake2013}
S~Alben.
\newblock {Optimizing snake locomotion in the plane}.
\newblock {\em Proc. Roy. Soc. A}, 469(2159):1--28, 2013.

\bibitem{taylor1952analysis}
Geoffrey~Ingram Taylor.
\newblock Analysis of the swimming of long and narrow animals.
\newblock {\em Proc. R. Soc. Lond. A}, 214(1117):158--183, 1952.

\bibitem{sfakiotakis2009undulatory}
Michael Sfakiotakis and Dimitris~P Tsakiris.
\newblock {Undulatory and pedundulatory robotic locomotion via direct and
  retrograde body waves}.
\newblock In {\em {Robotics and Automation, 2009. ICRA'09. IEEE International
  Conference on}}, pages 3457--3463. IEEE, 2009.

\bibitem{maladen2011undulatory}
Ryan~D Maladen, Yang Ding, Paul~B Umbanhowar, and Daniel~I Goldman.
\newblock Undulatory swimming in sand: experimental and simulation studies of a
  robotic sandfish.
\newblock {\em The International Journal of Robotics Research}, 30(7):793--805,
  2011.

\bibitem{JiAl2013}
F~Jing and S~Alben.
\newblock {Optimization of two- and three-link snake-like locomotion}.
\newblock {\em Physical Review E}, 87(2):022711, 2013.

\bibitem{cox1970motion}
RG~Cox.
\newblock {The motion of long slender bodies in a viscous fluid. Part 1.
  General theory}.
\newblock {\em Journal of Fluid Mechanics}, 44(04):791--810, 1970.

\bibitem{wang2014optimizing}
Xiaolin Wang, Matthew~T Osborne, and Silas Alben.
\newblock Optimizing snake locomotion on an inclined plane.
\newblock {\em Physical Review E}, 89(1):012717, 2014.

\bibitem{wang2018dynamics}
Xiaolin Wang and Silas Alben.
\newblock Dynamics and locomotion of flexible foils in a frictional
  environment.
\newblock {\em Proc. R. Soc. A}, 474(2209):20170503, 2018.

\bibitem{bhushan2013introduction}
Bharat Bhushan.
\newblock {\em Introduction to tribology}.
\newblock John Wiley \& Sons, 2013.

\bibitem{popov2010contact}
Valentin~L Popov.
\newblock {\em Contact mechanics and friction}.
\newblock Springer, 2010.

\bibitem{pennestri2016review}
Ettore Pennestr{\`\i}, Valerio Rossi, Pietro Salvini, and Pier~Paolo Valentini.
\newblock Review and comparison of dry friction force models.
\newblock {\em Nonlinear dynamics}, 83(4):1785--1801, 2016.

\bibitem{kelley1995iterative}
CT~Kelley.
\newblock {\em Iterative methods for linear and nonlinear equations}.
\newblock Society for Industrial and Applied Mathematics, 1995.

\bibitem{dennis1996numerical}
JE~Dennis~Jr and Robert~B Schnabel.
\newblock {\em Numerical Methods for Unconstrained Optimization and Nonlinear
  Equations}, volume~16.
\newblock SIAM, 1996.

\bibitem{hopkins2009survey}
James~K Hopkins, Brent~W Spranklin, and Satyandra~K Gupta.
\newblock {A survey of snake-inspired robot designs}.
\newblock {\em Bioinspiration \& Biomimetics}, 4(2):021001, 2009.

\bibitem{Pu1977a}
E~M Purcell.
\newblock {Life at low Reynolds number}.
\newblock {\em Am. J. Phys}, 45(1):3--11, 1977.

\bibitem{AvRa2008a}
J~E Avron and O~Raz.
\newblock {A geometric theory of swimming: Purcell's swimmer and its
  symmetrized cousin}.
\newblock {\em New Journal of Physics}, 10:063016, 2008.

\bibitem{KaMaRoMe2005a}
E~Kanso, J~E Marsden, C~W Rowley, and J~B Melli-Huber.
\newblock {Locomotion of articulated bodies in a perfect fluid}.
\newblock {\em Journal of Nonlinear Science}, 15(4):255--289, 2005.

\bibitem{HaCh2010a}
R~L Hatton and H~Choset.
\newblock {Connection vector fields and optimized coordinates for swimming
  systems at low and high Reynolds numbers}.
\newblock In {\em Proceedings of the ASME Dynamic Systems and Controls
  Conference (DSCC), Cambridge, Massachusetts, USA}, 2010.

\bibitem{jayne1991kinematics}
Bruce~C Jayne and James~D Davis.
\newblock Kinematics and performance capacity for the concertina locomotion of
  a snake (coluber constrictor).
\newblock {\em Journal of Experimental Biology}, 156(1):539--556, 1991.

\bibitem{keller1983crawling}
Joseph~B Keller and Meira~S Falkovitz.
\newblock Crawling of worms.
\newblock {\em Journal of Theoretical Biology}, 104(3):417--442, 1983.

\end{thebibliography}

\end{document}